\documentclass[fleqn,usenatbib]{mnras}

\usepackage{newtxtext,newtxmath}

\usepackage[T1]{fontenc}

\DeclareRobustCommand{\VAN}[3]{#2}
\let\VANthebibliography\thebibliography
\def\thebibliography{\DeclareRobustCommand{\VAN}[3]{##3}\VANthebibliography}

\usepackage{graphicx,rotating}
\usepackage{amsmath,pdflscape}
\usepackage{natbib}
\usepackage{subeqnarray}
\usepackage{empheq}
\usepackage{mathtools}
\usepackage{cases}
\usepackage{upgreek}
\usepackage{tikz}

\newcommand{\hatrho}{{\hat{\rho}}}
\newcommand{\hatomega}{{\hat{\Omega}}}
\newcommand{\du}{{\rm d}}

\newcommand{\bare}{\bar{\epsilon}}

\newcommand{\llbra}{[\mkern-2.5mu[}
\newcommand{\rrbra}{]\mkern-2.5mu]}
\defcitealias{h2022a}{Paper I}
\defcitealias{h2022b}{Paper II}
\defcitealias{ch33}{Chandra.(1933)}

\begin{document}

\title[On heterogenous configurations]{Nested spheroidal figures of equilibrium \\ IV. On heterogeneous configurations} 
\author[C. Staelen and J.-M. Hur{\'e}]
       {C. Staelen$^{1}$\thanks{E-mail:clement.staelen@u-bordeaux.fr} and J.-M. Hur\'e$^{1}$\\
$^{1}$Univ. Bordeaux, CNRS, LAB, UMR 5804, F-33600 Pessac, France}

\date{Accepted 2023 October 16. Received 2023 September 19; in original form 2023 June 7}
 
\pagerange{\pageref{firstpage}--\pageref{lastpage}} \pubyear{2023}

\makeatletter
\g@addto@macro\bfseries{\boldmath}
\makeatother
\maketitle

\label{firstpage}

\begin{abstract}
The theory of Nested Figures of Equilibrium, expanded in Papers I \& II, is investigated in the limit where the number of layers of the rotating body is infinite, enabling to reach full heterogeneity. In the asymptotic process, the discrete set of equations becomes a differential equation for the rotation rate. In the special case of rigid rotation (from center to surface), we are led to an Integro-Differential Equation (IDE) linking the ellipticity of isopycnic surfaces to the equatorial mass-density profile. In constrast with most studies, these equations are not restricted to small flattenings, but are valid for fast rotators as well. We use numerical solutions obtained from the SCF-method to validate this approach. At small ellipticities (slow rotation), we fully recover Clairaut's equation. Comparisons with Chandrasekhar's perturbative approach and with Roberts' work based on Virial equations are successful. We derive a criterion to characterize the transition from slow to fast rotators. The treatment of heterogeneous structures containing mass-density jumps is proposed through a modified IDE.
\end{abstract}

\section{Introduction}

Unveiling the internal structure of celestial bodies is a longstanding and fundamental challenge in astrophysics. Theories have emerged three centuries ago, with a principal interest in the Earth's interior. In the limit of slow rotation, \citet{clairaut43} showed the isopycnic surfaces are spheroids, i.e. ellipsoids of revolution. Using the theory of \citet{maclaurin42} for homogeneous spheroids, he obtained a second-order, ordinary differential equation linking the flattening of isopycnics to the mass-density profile. This equation has been more recently extended by \citet{lan62,lan74}, the shape of the external surface being expanded over Legendre polynomials  $P_{2n}$ up to the $n$-th order. Unfortunately, Clairaut's equation admits essentially no analytical solutions \citep[with some exceptions, see][]{tisserand91,mar00}. Slow rotators are accessible from the ``modified'' Lane-Emden equation in the form of series \citep{ch33,kov68}.  Besides, Clairaut's equation is limited to small flattenings  (i.e. to low rotation rates), while many systems do not belong to the category of slow rotators. This is the case of the giant planets in the Solar System. For Jupiter and Ceres, the flattening parameter $f \approx 0.07$, and this is even larger for Saturn \citep{tri14,rcc15}. Achernar represents an extreme configuration \citep{cado2008}. New developments remain therefore necessary to model the structure of spinning objects, especially for moderate to fast rotation rates \citep[e.g.][]{lan62,ragazzo2020}.

The determination of the gravitational potential  of rotating bodies has always demanded a high analytical effort or substantial computational resources, or both (this exceeds the present context). The spheroidal shape is appealing, as its gravitational potential is known with a closed form \citep[see e.g.][]{chandra69}.  \citet{kzs15} have investigated the validity of the hypothesis of spheroidal isopycnics by comparing of the ``true'' shape obtained by numerical means to ``perfect'' spheroids. They showed that discrepancies are small in amplitude. In fact, this remains true at moderate/large rotation, but unsurprisingly fails close to the mass-shedding limit \citep{hachisu86}.  Using the gravitational potential of a heterogeneous spheroid, \citet{rob63} used the tensor Virial theorem to derive equations valid for fast rotators, but no self-consistent solutions was produced.

In \citet[][hereafter, Paper I and II, respectively]{h2022a,h2022b}, we have investigated the conditions of equilibrium of a piece-wise, heterogeneous system made of $\cal L$ homogeneous layers bounded by spheroidal surfaces. The theory of Nested Spheroidal Figures of Equilibrium (hereafter, NSFoE) assumes that these surfaces stay close to confocality (in the sense of oblate spheroidal coordinates; see Sect. \ref{sec:id}) and that each layer can rotate at its own rate \citep{veronet12,bbm15}. A wide range of configurations is then reachable, from quasi-spheres to very flat, disk-like objects. It must be pointed out that such solutions remain approximate, although the Virial parameter relative to the gravitational energy is very small \citep{sta22}. This is a consequence of Poincar\'e-Hamy theorems: a rigidly rotating body with a spheroidal stratification is not an exact figure of equilibrium\footnote{Only confocal surfaces can lead to an exact equilibrium when all layers rotate in a synchroneous manner. This equilibrium requires a mass-density inversion, which, for stability reasons, is physically not tenable \citep{poincare88,hamy90,1903volterra}.}. In this article, we investigate the solutions in the case where the number ${\cal L}$ of layers is infinite, which corresponds to a fully heterogeneous body, and for a global rigid rotation. It is therefore a natural continuation of Paper II. Another motiviation of the article is the case of moderate/fast rotators, characterized by a significant oblateness or flatenning (larger than a percent typically). It is therefore interesting to see to what extent classical theories, which are often limited to slow rotation, remain valid or fail. In this purpose, it is necessary to compare any analytical result with numerical solutions. In the present case, this is achieved by using the  {\tt DROP} code which solves the problem for a polytopic equation-of-state (EoS), various flattenings and rotation profiles \citep{hh17,bh21}.
 
It is obvious that the present approach is not supposed to surpass sophisticated models for stars and planets, which are dynamically and thermodynamically more complex that what the hypothesis made here allow. Stars are widely prone to mixing, transport and circulation. Planets, closer to rigid rotation, have a more simple layered structure (except at the very surface) and isopynics surfaces are believed to be very close to spheroids, as suggested by the inversion of gravitational moments \citep[e.g.][]{hub13,net21}. After a brief summary on the theory NSFoE, we show in Sec. \ref{sec:id} how the discrete set of equations can be converted into a differential equation. For rigid, global rotation, this is equivalent to an Integro-Differential Equation (IDE) for the ellipticity of isopycnics. In a first example, we feed this IDE with the numerical solutions obtained from the Self-Consistent-Field (SCF) method \citep[e.g.][]{hachisu86}, and show that this approach is not only coherent but quite accurate. In Sec. \ref{sec:clairaut}, we study the behavior of the equation in the limit of small flattenings, which happens at slow rotation. In particular, we show that the formalism is fully compatible with classical theories, namely the fundamental second-order differential equation established by \citet{clairaut43}, the solutions obtained by \citet{ch33} from the ``modified'' Lane-Emden equation. We also make a comparison with the equation of \citet{rob63}. The question of internal jumps is adressed in Sec. \ref{sec:jumps}, where we derive a modified IDE and test it. In the concluding section, we propose a criterion characterizing the transition from slow to fast rotators, and give a few perspectives.

\begin{figure}
       \centering
       \includegraphics[width=.8\linewidth]{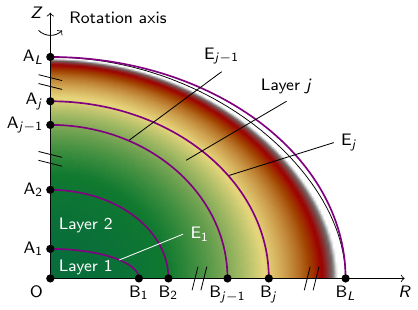}
       \caption{Typical configuration for a heterogeneous body with finite number of homogeneous layers (${\cal L}$ in total) bounded by spheroidal surfaces ${\rm E}_j$.
       }
       \label{fig:system}
\end{figure}

\section{Theory for heterogeneous bodies}
\label{sec:id}

\subsection{Equation set for the theory of NSFoE}

We adopt the same theoretical background and same notations as in \citetalias{h2022a} and \citetalias{h2022b}, which can be summarized as follows. We consider $\cal L$ oblate, non-intersecting spheroidal surfaces ${{\rm E}_j, \; j\in\llbra1,{\cal L}\rrbra}$ with semi-major axis $a_j$, semi-minor axis $0 < b_j \leq a_j$ and eccentricity
\begin{equation}\label{eq:def_epsilonj}
  \epsilon_j=\sqrt{1-b^2_j/a_j^2},
\end{equation}
as depicted in Fig. \ref{fig:system}. These surfaces define $\cal L$ layers. We note with index $j \ge 2$ the layer bounded by ${\rm E}_{j-1}$ and ${\rm E}_j$ (we then have ${a_{j-1}<a_j}$ and ${b_{j-1}<b_j}$). Index $1$ corresponds to the deepest layer, bounded by surface $\rm E_1$ only. Furthermore, we assume that each layer $j$ is homogeneous, with mass density $\rho_j$, and rotates rigidly around the $Z$-axis at a rate $\Omega_j$. A key point in the theory of NFSoE is the possibility of asynchroneous motion of layers, i.e. $\Omega_{j-1} \neq \Omega_j$. In this article, however, we will consider a subclass of configurations characterized by synchroneous rotations.

A fundamental parameter that controls the applicability of the theory is the ``confocal parameter'' $c_{i,j}$, defined for each pair  $({\rm E}_i,{\rm E}_j)$ by
\begin{equation}\label{eq:def_cij}
       c_{i,j} = q_{i,j}^2\epsilon_i^2-\epsilon_j^2.
\end{equation}
This parameter is positive if a surface ${\rm E}_i$, interior to a surface ${\rm E}_j$ is, in terms of oblate spheroidal coordinates, more oblate than layer $j$. As quoted in the introduction, only systems with confocal spheroidal surfaces (i.e. $ c_{i,j}=0$ for all pairs) correspond to an exact equilibrium \citep{poincare88,hamy90}. Then, the equilibrium of any layered systems in rigid rotation with non-zero confocal parameters is necessarily approximate. As shown in \citetalias{h2022a} and \citetalias{h2022b}, equilibria with $|c_{i,j}|\lesssim0.3$ typically are found to be very close to numerical simulations obtained with the {\tt DROP}-code \citep{hh17,bh21}. Besides, the $c$-parameter is generally found to be slightly negative, meaning that isopycnic surfaces tend to be more spherical with depth in the system, or equivalently, that the ellipticity of isopycnics increases from the center to the surface. However, note that if models of stars and planets mainly agree on such a ``standard'' stratification, there is no argument or observational proof that definitively rules out a reversal, for some objects. This may depend on physical mechanisms at work and on the formation process. Prolate shapes can be induced by circulations or magnetic fields \cite[e.g.][and references therein]{fe14}.

The starting point of the present work is (27) of \citetalias{h2022b}, which links the properties of all layers together. This is a set of coupled, ${\cal L}-1$ algebraic equations, which read,
\begin{align}\label{eq:om2_nsfoe}
       &\frac{\alpha_j \tilde\Omega_j^2-\tilde\Omega_{j+1}^2}{\alpha_j-1} =\\
       &\mkern+10mu \sum_{i=1}^{j-1} \tilde\rho_{i+1}(\alpha_i-1)\frac{\bare_i}{\epsilon_i^3}\bigg[2\arcsin\left(\frac{q_{i,j}\epsilon_i}{\sqrt{1+c_{i,j}}}\right)(1+c_{i,j})\notag\\
       &\mkern+5mu +(1-2q_{i,j}^2\epsilon_i^2)\arcsin(q_{i,j}\epsilon_i) -q_{i,j}\epsilon_i\sqrt{1-q_{i,j}^2\epsilon_i^2} -2q_{i,j}\epsilon_i\bare_j \bigg]\notag \\
       &\mkern+5mu +\sum_{i=j}^{\cal L} \tilde\rho_{i+1}(\alpha_i-1)\bigg[{\cal M}(\epsilon_i)+\frac{2}{\epsilon_i^2}(\epsilon_j^2-\epsilon_i^2)\bigg] \notag
\end{align}
for $j < {\cal L}$, where
\begin{equation}
       \left\{
              \begin{aligned}
                     &\tilde\rho_j = \rho_j/\rho_{\cal L},\\
                     &\alpha_j = \rho_j/\rho_{j+1},\\
                     &q_{i,j} = a_i/a_j,\\
                     &\bare_j = b_j/a_j,
              \end{aligned}
       \right.
\end{equation}
and
\begin{equation}
       {\cal M}(\epsilon)=\left(3-2\epsilon^2\right)\frac{\bar{\epsilon}}{\epsilon^3}\arcsin(\epsilon)+3-\frac{3}{\epsilon^2}\label{eq:mlfunction},
\end{equation}
 is Maclaurin's function defined by (see \citetalias{h2022a}), and
\begin{equation}
       \tilde\Omega_j = \frac{\Omega_j}{\sqrt{2\uppi G\rho_{\cal L}}}.
\end{equation}
is the dimensionless rotation rate normalised to the mass density of the uppermost layer ($G$ is the gravitational constant). For $j={\cal L}$ (the upper layer), we have
\begin{align}\label{eq:om2_nsfoe_ext}
       \tilde\Omega_{\cal L}^2= &\sum_{i=1}^{\cal L} \tilde\rho_{i+1}(\alpha_i-1)\frac{\bare_i}{\epsilon_i^3}\bigg[2\arcsin\left(\frac{q_{i,{\cal L}}\epsilon_i}{\sqrt{1+c_{i,{\cal L}}}}\right)(1+c_{i,{\cal L}})\notag\\
       &\quad+(1-2q_{i,{\cal L}}^2\epsilon_i^2)\arcsin(q_{i,{\cal L}}\epsilon_i)-q_{i,{\cal L}}\epsilon_i\sqrt{1-q_{i,{\cal L}}^2\epsilon_i^2} \notag\\
       &\qquad-2q_{i,{\cal L}}\epsilon_i\bare_{\cal L} \bigg].
\end{align}

\subsection{From a discrete set of layers to a continuum}

We now seek for equilibrium configurations where $\rho$ is continuous and derivable from the center to the surface. We first consider configurations without any mass-density jumps (mass-density jumps are considered in Sec. \ref{sec:conclusion}). As each layer $j$ has its specific mass-density, and specific set of confocal parameters $c_{i,j}$, the theory of NSFoE is expected to be capable of such a prolongation, provided these confocal parameters are all ``small'' enough. When ${\cal L}$ drastically increases, the extension of layer $j$ in the equatorial plane is $a_{j}-a_{j-1}\equiv\Delta a_j \rightarrow 0$. In a similar way, at the polar axis, we have $b_j-b_{j-1}\equiv\Delta b_j\rightarrow 0$. Furthermore, the difference in the mass-density between two consecutive layers is
\begin{align}
       \Delta \rho_j = \rho_{j+1}-\rho_{j} &= \rho_{\cal L}\Delta\tilde\rho_j \notag\\
       &=\rho_{\cal L}\tilde\rho_{j+1}(\alpha_j-1)\rightarrow 0.
\end{align}
In these conditions, \eqref{eq:om2_nsfoe} can be rewritten as
\begin{align}
       &-\frac{\tilde\rho_{j+1}\tilde\Omega_{j+1}^2-\tilde\rho_j\tilde\Omega_j^2}{\Delta\tilde\rho_j} = \notag\\
       &\mkern+25mu \sum_{i=1}^{j-1} \Delta\tilde\rho_i\frac{\bare_i}{\epsilon_i^3}\bigg[2\arcsin\left(\frac{q_{i,j}\epsilon_i}{\sqrt{1+c_{i,j}}}\right)(1+c_{i,j}) -2q_{i,j}\epsilon_i\bare_j \notag\\
       &\mkern+75mu +(1-2q_{i,j}^2\epsilon_i^2)\arcsin(q_{i,j}\epsilon_i) -q_{i,j}\epsilon_i\sqrt{1-q_{i,j}^2\epsilon_i^2}\bigg]\notag\\
       &\mkern+40mu +\sum_{i=j}^{\cal L} \Delta\tilde\rho_i\bigg[{\cal M}(\epsilon_i)+\frac{2}{\epsilon_i^2}(\epsilon_j^2-\epsilon_i^2)\bigg].\label{eq:om2_delta}
\end{align}
In this form, we see that \eqref{eq:om2_delta} has the convenient form for the continuous case, as in the limit $\Delta\tilde\rho_i \rightarrow 0$ the sums over $i$ tend to integrals. To express these integrals, the equatorial radius of layer $j$  is rewritten in the form of the dimensionless, continuous variable
\begin{equation}
       \varpi \equiv \frac{a_j}{R_{\rm e}}.
\end{equation}
Equivalentally, this is the semi-major axis of the isopycnic surface ${\rm E}(\varpi)\equiv {\rm E}_j$, normalised to the equatorial radius of the body. In a similar manner, we associate
\begin{equation}
       \varpi' \equiv \frac{a_i}{R_{\rm e}},
\end{equation}
with the equatorial radius of layer $i$. As long as $\epsilon_1\neq1$, we have ${a_1\rightarrow0}$, otherwise a minimal radius is required. Yet, numerical solutions obtained from the {\tt DROP}-code (see Figs. \ref{fig:configA}-\ref{fig:configD}) show that the deeper the layer, the smaller its flattening, namely ${\nabla\epsilon>0}$, in agreement with classical theories. Thus, in this work, we will freely take ${a_1\rightarrow0}$, so that ${(\varpi',\varpi)\in[0,1]^2}$.

In the perspective of a continuous mass-density profile, we must consider situations where the mass density vanishes onto the external surface. In general, the \emph{adimensionning used for the discrete theory is not appropriate} and must be reconsidered. This is easily corrected. In this purpose, we choose the \emph{central} mass density ${\rho_{\rm c}=\rho_1}$ as the new reference, instead of $\rho_{\cal L}$. The main parameters of layer $j$ are then
\begin{equation}
       \left\{
              \begin{aligned}
                     &\hatrho(\varpi) \equiv \frac{\rho_j}{\rho_{\rm c}} \equiv \tilde\rho_j \times \frac{\rho_{\cal L}}{\rho_1},\\
                     &\hatomega(\varpi) \equiv \frac{\Omega_j}{\sqrt{2\uppi G\rho_{\rm c}}} \equiv \tilde\Omega_j \times \sqrt{\frac{\rho_{\cal L}}{\rho_1}},\\
                     &\epsilon(\varpi) \equiv \epsilon_j,\\
                     &\bare(\varpi)= \sqrt{1-\epsilon^2(\varpi)} \equiv \bare_j,
              \end{aligned}
       \right.
\end{equation}
where $\epsilon(\varpi)$ is the eccentricity of the isopycnic surface ${\rm E}(\varpi)$ and $\bare(\varpi)$ is its axis ratio. In a similar way, the confocal parameter is now a continuous variable, namely, from (\ref{eq:def_cij})
\begin{equation}\label{eq:confoc_def}
       c(\varpi',\varpi) = \frac{\varpi'^2}{\varpi^2}\epsilon^2(\varpi')-\epsilon^2(\varpi). \mkern+25mu \big(\equiv c_{i,j}\big)
\end{equation}

\begin{figure}
       \centering
       \includegraphics[trim={2.5mm 0 2.1mm 0},clip,width=\linewidth]{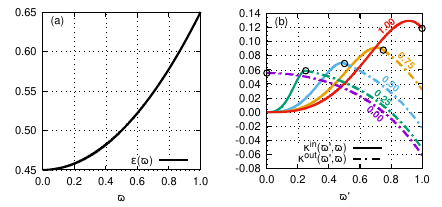}
       \caption{Eccentricity profile ({\it left}) prescribed for the example. Functions $\kappa^{\rm in}$ and $\kappa^{\rm out}$ ({\it right}), defined by \eqref{eq:kin_def} and \eqref{eq:kout_def} respectively, versus $\varpi'$ for $\varpi\in\{0.00,0.25,0.50,0.75,1.00\}$ (labelled along the curves). Voided circles pin the value of the functions on $\varpi'=\varpi$.}
       \label{fig:kappa}
\end{figure}

With these definitions, the left-hand side (LHS) of \eqref{eq:om2_delta} becomes 
\begin{equation}\label{eq:lhs}
       \frac{\tilde\rho_{j+1}\tilde\Omega_{j+1}^2-\tilde\rho_j\tilde\Omega_j^2}{\Delta\tilde\rho_j} \rightarrow \frac{\du(\hatrho\hatomega^2)}{\du\hatrho},
\end{equation}
while the right-hand side (RHS) of \eqref{eq:om2_delta}, more complex, can be written in compact form as
\begin{align}\label{eq:rhs}
  \text{RHS of \eqref{eq:om2_delta}}& \rightarrow \int_{\hatrho(0)}^{\hatrho(\varpi)} d\hatrho(\varpi') \kappa^{\rm in}(\varpi',\varpi)\\
  & \mkern+75mu  + \int_{\hatrho(\varpi)}^{\hatrho(1)} d\hatrho(\varpi') \kappa^{\rm out}(\varpi',\varpi), \notag
\end{align}
where the two functions $\kappa^{\rm in}$ and $\kappa^{\rm out}$ are explicitely given in Appendix \ref{sec:kappa_chi_mu}; see (\ref{eq:kin_def}) and (\ref{eq:kout_def}). Formally, these depend on $\varpi$, $\varpi'$, and $\epsilon$, i.e. $\kappa \equiv \kappa(\varpi',\varpi; \epsilon)$. As $\epsilon$ depends on $\varpi'$ or $\varpi$, then there are only two variables on input. Despite apparences, these functions behave very well over the integration range. Among interesting properties, we have ${\kappa^{\rm in}(\varpi,\varpi)=\kappa^{\rm out}(\varpi,\varpi)}$. This is particularly important and attractive for numerical applications. We also see that $\kappa^{\rm in}(\varpi,0)=0$ for $\varpi\neq0$ ($\kappa^{\rm in}$ is not defined for this value). The typical shape of the these functions is visible in  Fig. \ref{fig:kappa}b, where we have plotted $\kappa^{\rm in}$ and $\kappa^{\rm out}$ as functions of $\varpi'$ for fives values of $\varpi$. For this exemple, we have prescribed a parabolic profile for the eccentricity (see Fig. \ref{fig:kappa}a), as observed in many numerical experiments (see, e.g. configuration A discussed below). 

It follows from \eqref{eq:lhs} and \eqref{eq:rhs} that \eqref{eq:om2_delta} reads, in the continuous limit
\begin{align}\label{eq:drhoom2_nsfoe}
 -\frac{\du (\hatrho\hatomega^2)}{\du\varpi} = &\frac{\du\hatrho}{\du\varpi}\Bigg[\int_{\hatrho(0)}^{\hatrho(\varpi)} \du\hatrho(\varpi') \kappa^{\rm in}(\varpi',\varpi) \notag\\
 & \mkern+75mu + \int_{\hatrho(\varpi)}^{\hatrho(1)} \du\hatrho(\varpi') \kappa^{\rm out}(\varpi',\varpi)\Bigg].
\end{align}
This equation is the \emph{main equation of the present problem}. It links the eccentricity of the isopycnic surfaces and their mass density (in fact, its derivative) to the variations of the rotation rate. It enables to reach configurations where both the mass density and rotation rate vary smoothly with the equatorial radius. We can apply the same transformation to \eqref{eq:om2_nsfoe_ext}, and we obtain at the surface
\begin{equation}\label{eq:drhoom2_nsfoe_ext}
       -\hatomega^2(1) = \int_{\hatrho(0)}^{\hatrho(1)} \du\hatrho(\varpi') \kappa^{\rm in}(\varpi',1).
\end{equation}
We see that the RHS of \eqref{eq:drhoom2_nsfoe} and \eqref{eq:drhoom2_nsfoe_ext} coincide for $\varpi=1$, within a factor $(\du\hatrho/\du\varpi)|_{\varpi=1}$. So, the two LHS must also coincide for $\varpi=1$, which imposes the condition
\begin{equation}\label{eq:dom2_surf}
       \frac{\du\hatomega^2}{\du\varpi}\Bigg|_{\varpi=1} = 0.
\end{equation}
Thus, the squared rotation has an extremum at the surface.

\subsection{The case of global rigid rotation: the general Integro-Differential Equation (IDE) for the eccentricity}

\begin{figure}
       \centering
       \includegraphics[trim={2.5mm 0 2.1mm 0},clip,width=\linewidth]{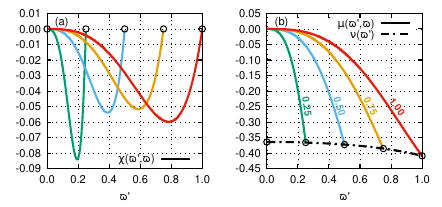}
       \caption{Function $\chi$ ({\it left panel}) defined by (\ref{eq:chi_def}), and functions $\mu$ and $\nu$  ({\it right panel}) defined by (\ref{eq:mu_def}) and (\ref{eq:nu_def}), respectively, versus $\varpi'$ for ${\varpi \in \{0.25, 0.50, 0.75, 1.00\}}$ (labelled along the curves). Voided circles pin the value of the functions on $\varpi'=\varpi$. The eccentricity profile used is the same as in Fig. \ref{fig:kappa}.}
       \label{fig:chi_mu}
\end{figure}

We see that (\ref{eq:drhoom2_nsfoe}) is capable of modeling a wide range of situations, from rigid to differential rotation, and independently, from homogeneous to heterogeneous mass-density profiles. In this work, we focus on rigidly rotating bodies, so we have $\du\hatomega=0$. It means that \eqref{eq:dom2_surf} is naturally satisfied. Therefore, \eqref{eq:drhoom2_nsfoe} becomes
\begin{align}\label{eq:om2_mono}
   \notag
   -\hatomega^2 = &\int_{\hatrho(0)}^{\hatrho(\varpi)} \du\hatrho(\varpi') \kappa^{\rm in}(\varpi',\varpi) \\
 &\mkern+75mu+\int_{\hatrho(\varpi)}^{\hatrho(1)} \du\hatrho(\varpi')\kappa^{\rm out}(\varpi',\varpi).
\end{align}
This equation yields the rotation rate of the body once the configuration is known through $\hatrho(\varpi)$ and $\epsilon(\varpi)$. Alternatively, it can be used to constrain the solutions if the rotation law is prescribed in advance. We can take the derivative of \eqref{eq:om2_mono} with respect to $\varpi$. In this purpose, we use Leibniz's integral rule, namely, for a given derivable function $g$,
\begin{equation}
       \frac{\du}{\du x}\int_{x_0}^x \du y \, g(x,y) = g(x,x) + \int_{x_0}^x \du y \, \frac{\partial}{\partial x}g(x,y),
\end{equation}
where $x_0$ is a constant. In the present case, it leads to
\begin{align}
 \int_{\hatrho(0)}^{\hatrho(\varpi)} \du \hatrho(\varpi') &\frac{\partial}{\partial \varpi }\kappa^{\rm in}(\varpi ',\varpi ) \notag\\
 & + \int_{\hatrho(\varpi)}^{\hatrho(1)} \du \hatrho(\varpi') \frac{\partial}{\partial \varpi}\kappa^{\rm out}(\varpi',\varpi) = 0,
  \label{eq:dide}
\end{align}
where we have used the property that $\kappa$ is continuous at $\varpi'=\varpi$ (see above). In fact, the partial derivatives can be put in the form

\begin{equation}
  \frac{\partial\kappa^{\rm in}}{\partial \varpi} = 4\chi(\varpi',\varpi) - 2\frac{\du\epsilon^2}{\du \varpi}\mu(\varpi',\varpi),
  \label{eq:dkin}
 \end{equation}
and
\begin{equation}
  \frac{\partial\kappa^{\rm out}}{\partial \varpi} = - 2\frac{\du\epsilon^2}{\du \varpi}\nu(\varpi'),
  \label{eq:dkout}
\end{equation}
where $\chi$, $\mu$ and $\nu$ are defined in the Appendix \ref{sec:kappa_chi_mu}; see (\ref{eq:chi_def}), (\ref{eq:mu_def}) and (\ref{eq:nu_def}), respectively. Like $\kappa$, these functions depend on $3$ quantities, $\varpi$, $\varpi'$ and $\epsilon$, but implicitly only on the $2$ space variables $\varpi$ and $\varpi'$. An illustration is given in Fig. \ref{fig:chi_mu} for the parabolic eccentricity profile considered previously. An important property is that $\chi(\varpi,\varpi)=0$ and $\mu(\varpi,\varpi)=\nu(\varpi)$, which means that the derivatives of the $\kappa$-functions are equal at the connection, i.e. 
\begin{equation}
       \left.\frac{\partial\kappa^{\rm in}}{\partial \varpi}\right|_{\varpi'=\varpi} = \left.\frac{\partial\kappa^{\rm out}}{\partial \varpi}\right|_{\varpi'=\varpi}.
\end{equation}
This is visible in Fig. \ref{fig:kappa}b. These functions have also a relatively small amplitude, which, again, is very practial for any numerical treatment. From \eqref{eq:dkin} and \eqref{eq:dkout}, \eqref{eq:dide} becomes
\begin{align}\label{eq:id_nsfoe}
      2\int_{\hatrho(0)}^{\hatrho(\varpi)} \du\hatrho(\varpi') \chi(\varpi',\varpi) = &\frac{\du\epsilon^2}{\du \varpi}\Bigg[\int_{\hatrho(0)}^{\hatrho(\varpi)} \du\hatrho(\varpi')\mu(\varpi',\varpi) \notag \\
      &+ \int_{\hatrho(\varpi)}^{\hatrho(1)}\du\hatrho( \varpi')\nu( \varpi')\Bigg] ,
\end{align}
which then links directly the eccentricity of the isopycnic surfaces to their mass density. 

A consequence of \eqref{eq:id_nsfoe} comes from the case $\varpi=0$ (i.e. the center of the body). Indeed, at this point, we have
\begin{equation}\label{eq:id_nsfoe_zero}
 \left.\frac{\du\epsilon^2}{\du \varpi}\right|_{\varpi=0}\times\int_{\hatrho(0)}^{\hatrho(1)}\du\hatrho(\varpi')\nu(\varpi') = 0.
\end{equation}
As $\nu(\varpi')<0 \; \forall \varpi'$ (this is seen from its definition; see \eqref{eq:nu_def} in the Appendix) and $\du\hatrho<0$ from stability consideration, \eqref{eq:id_nsfoe_zero} yields
\begin{equation}\label{eq:de2da0}
       \left. \frac{\du\epsilon^2}{\du \varpi}\right|_{\varpi=0} = 0,
\end{equation}
whatever the mass density profile.

\subsection{A note on the condition of immersion}

As early quoted in this section, an important hypothesis of the theory of NSFoE is the non-intersection of the interfaces between layers. In the continuous limit, this means that the isopycnic surfaces must not cross each other. Two ellipses intersect if and only if the one with the largest major axis has also the smallest minor axis. So, if the polar radius $b$, i.e. the minor axis, is given by $b(\varpi)/R_{\rm e}=\varpi\bare(\varpi)$, we require
\begin{equation}\label{eq:immersion_b}
       \frac{\du [\varpi\bare(\varpi)]}{\du \varpi} > 0.
\end{equation}

It is easily shown from the definitions of $\epsilon$ and $\bare$ that
\begin{equation}
       \frac{\du [\varpi\bare(\varpi)]}{\du \varpi} = \bare - \frac{\varpi}{2\bare}\frac{\du\epsilon^2}{\du \varpi}.
\end{equation}
So, \eqref{eq:immersion_b} can be written as 
\begin{equation}\label{eq:immersion_e2}
       \frac{\du\epsilon^2}{\du \varpi} < \frac2\varpi\left[1-\epsilon^2(\varpi)\right],
\end{equation}
which then imposes an \emph{upper limit for the eccentricity gradient}. Note that requiring ${\varpi(\du\epsilon^2/\du \varpi) = 2[1-\epsilon^2(\varpi)]}$ ${\forall \varpi\in[0,1]}$ leads to $\epsilon^2(\varpi)=1$, i.e. the body would be infinitely flat.

\subsection{The particular case of homogeneity: Maclaurin formula recovered}
\label{ssec:maclaurin}

A first check of the \eqref{eq:id_nsfoe} is performed by considering the Maclaurin spheroid. In this case $\du\hatrho=0$, and the mass density profile is
\begin{equation}
       \hatrho(\varpi) = {\cal H}(1-\varpi),
\end{equation}
where $\cal H$ is Heaviside's step function. In the sense of distributions, the derivative of the mass density is
\begin{equation}
       \frac{\du\hatrho}{\du \varpi} = -\delta(1-\varpi),
\end{equation}
where $\delta$ is Dirac distribution. We can now use \eqref{eq:id_nsfoe} to obtain $\epsilon$, which is the only unknown of the problem. However, for a body where the mass density is a constant, the notion of isopycnic surfaces appears as a non-sense. Yet, the Poincar{\'e}-Wavre theorem implies that for a body where the rotation rate is constant on cylinders, isopycnic and isobaric surfaces must coincide\footnote{In fact, the theorem states equivalency between four propositions, two of which are used in this discussion; see e.g. \citet{tassoul78}.}. Rigid rotation has a rate which is obviously constant on cylinders, so we can consider $\epsilon$ as the eccentricity of isobaric surfaces. It can be shown that \eqref{eq:drhoom2_nsfoe} and \eqref{eq:id_nsfoe} become
\begin{align}
       \hatomega^2 = \kappa^{\rm out}(1,\varpi), \label{eq:om2_ml}
\end{align}
and
\begin{align}
\frac{\du\epsilon^2}{\du \varpi}\times\nu(1) = 0, \label{eq:de2da_ml}
\end{align}
respectively, where we used the properties of the Dirac distribution. From  \eqref{eq:nu_def}, we see that $\nu$ never vanishes, so the only solution to \eqref{eq:de2da_ml} is 
\begin{equation}\label{eq:de2da_ml_sol}
       \frac{\du\epsilon^2}{\du \varpi} = 0 \implies \epsilon^2(\varpi)=\epsilon^2(1),
\end{equation}
which means that isobaric surfaces are similar spheroids. So, by expliciting $\kappa^{\rm out}$, \eqref{eq:om2_ml} becomes 
\begin{equation}\label{eq:om2_ml_sol}
       \hatomega^2 = \left(3-2\epsilon_{\rm s}^2\right)\frac{\bare_{\rm s}}{\epsilon_{\rm s}^3}\arcsin(\epsilon_{\rm s}) + 3 - \frac{3}{\epsilon_{\rm s}^2} \equiv {\cal M}(\epsilon_{\rm s}),
\end{equation}
where $\epsilon_{\rm s} \equiv \epsilon(1)$ and $\bare_{\rm s}\equiv\bare(1)$ are values at the surface. As expected, we fully recover the results from Maclaurin's theory.

\subsection{Checking the IDE from a numerical reference}
\label{subsect:IDEcheck}

Unfortunately, without any prior knowledge on the mass-density profile or the eccentricity, \eqref{eq:id_nsfoe} can not solely be used to determine any internal structure. However, we can test the reliability of the above approach. In this purpose, we find more practical to rewrite \eqref{eq:id_nsfoe} in the form
\begin{align}\label{eq:id_nsfoe_bis}
\frac{\du\epsilon^2}{\du  \varpi} = \frac{2\int_{\hatrho(0)}^{\hatrho(\varpi)} \du\hatrho( \varpi') \chi( \varpi', \varpi)}{\int_{\hatrho(0)}^{\hatrho(\varpi)} \du\hatrho( \varpi')\mu( \varpi', \varpi)+ \int_{\hatrho(\varpi)}^{\hatrho(1)}\du\hatrho( \varpi')\nu( \varpi')}.
\end{align}
The test then consists in computing both sides of this expression by solutions obtained numerically from a Self-Consistent Field (SCF)-method. We use the {\tt DROP} code \citep{hh17,bh21} as {\it the numerical reference}. This code, which has been extensively used, solves the full structure of rotating, self-gravitating fluids for a wide range of flattenings, equation of states and rotation profiles\footnote{While the classical version of the {\tt DROP} code is typically second-order accurate in the mesh spacing, we have build an alternative version based spectral methods. This enables to reach much high precision (in a shorter computing time).}. A fundamental ingredient is the closure relationship between pressure $p$ and mass density $\rho$. In the paper throughtout, we use a polytropic EoS, namely
\begin{equation}
       p = K\rho^{1+1/n},
\end{equation}
where $K$ and $n$ (the polytropic index) are positive constants. Once the SCF-cycle has converged, the mass-density $\rho(R,Z)$ is known. Then, we have to determine isopycnic surfaces, denoted ${\rm S}_j$, from center to surface. Clearly, isopycnics are not exact spheroids (in general, these are sligthly depressed in the middle), but any equilibrium surface ${\rm S}_j$ crosses the polar axis and equatorial axis respectively at points ${\rm A}_j(0,b_j)$ and ${\rm B}_j(a_j,0)$ (see Fig. \ref{fig:system}). From these two points, we can calculate a ``pseudo-eccentricity'', basically from (\ref{eq:def_epsilonj}). This pseudo-eccentricity is then of the form $\epsilon(\varpi)$. We can then use this output, together with the mass-density $\rho(\varpi)$ along the equatorial plane to compute $\chi$, $\mu$, $\nu$ and the derivative of the pseudo-eccentricity $\du\epsilon^2/\du \varpi$, and then check (\ref{eq:id_nsfoe_bis}). We will also compare our results to Clairaut's integral equation, i.e. \eqref{eq:id_clrt}, which will be discussed in Sec. \ref{sec:clairaut}, and to Roberts' result,which is given in our notations in Appendix \ref{sec:rob} ; see \eqref{eq:roberts}.

There are four main sources of errors in this kind of numerical test. First, {\tt DROP} releases numerical solutions whose accuracy depends on the resolution. Second, the determination of equilibrium surface ${\rm S}_j$ (and then, points ${\rm A}_j$ and ${\rm B}_j$) is also not perfect. Next, the integrals in the RHS of (\ref{eq:id_nsfoe_bis}) are also sensitive to the quadrature scheme, as well as the scheme for the derivative of the pseudo-eccentricities (here, we use $2$nd-order schemes). Obviously, we do not expect \eqref{eq:id_nsfoe_bis} to be exactly satisfied. In turn, if both sides of this equation are very close for a broad variety of configuration, then it proves the reliability of the IDE.

\subsection{An example}
\label{subsect:example}

\begin{figure}
       \centering
       \includegraphics[trim={1mm 4.1cm 10mm 0},clip,width=\linewidth]{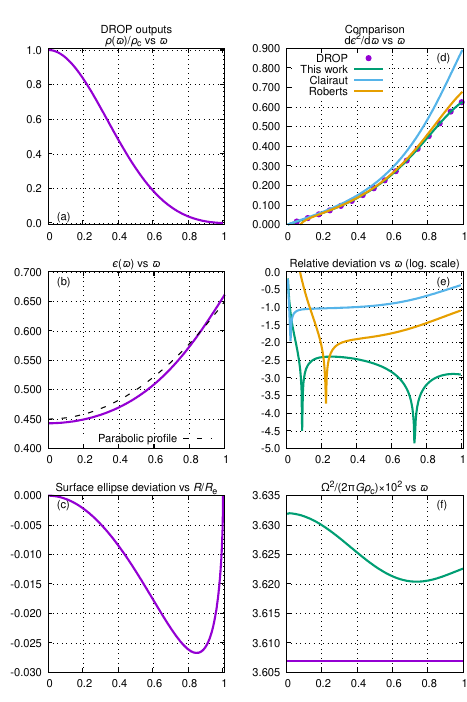}
       \caption{Results for configuration A ($\bare_{\rm s}=0.75$, $n=1.5$); see Tab. \ref{tab:configA} for global quantities. Left-hand side panels: outputs from the {\tt DROP} code, i.e. (a) eccentricity of the isopycnic layers (the dashed thin black line is the quadratic profile used for Figs. \ref{fig:kappa} and \ref{fig:chi_mu}); (b) radial mass density; (c) the deviation of the external surface from a spheroid. Right-hand side panels: Comparison between this work and the output from {\tt DROP}, i.e. (d) $\du\epsilon^2/\du \varpi$ as a function of $\varpi$; (e) decimal logarithm of the gap between the analytical methods and the numerical reference for $\du\epsilon^2/\du \varpi$; (f) decimal logarithm of the gap between {\tt DROP} and this work for the rotation rate.}
       \label{fig:configA}
\end{figure}

As a first illustration, we consider a rotating polytrope with surface axis-ratio $\bare_{\rm s}=0.75$ and polytropic index $n=1.5$, hereafter Configuration A. It corresponds to a fast rotator \citep[for comparison, Achernar has an axis ratio in surface around $0.74$; see][]{dkm14}, which is also one of the structures given in the tables of \citet{hachisu86}. The mass-density, pseudo-eccentricity and the deviation of the outermost surface to an exact spheroid are displayed in Fig. \ref{fig:configA}a to c (left panels). The RHS and LHS of \eqref{eq:id_nsfoe_bis} are plotted versus $\varpi$ in \ref{fig:configA} d. We see that the absolute deviation between these two estimates (panel e) is much less than $1\%$ for most radii, and even mess than $0.1 \%$ in the outerpart of the body. This agreement is already remarkable as the conforcal parameters (center and surface values) are marginally acceptable (i.e. $c(\varpi,1)\in[-0.4375,0]$). The figure also shows that the approximation is also valid for $\du\epsilon^2/\du \varpi$, as the discrepancy with {\tt DROP} is also of order $\sim10^{-3}$ in this case. From panel f, we see that the rotation rate $\hat\Omega$ deduced from \eqref{eq:om2_mono} is not strictly a constant, as would be expected. But, we see that it varies weakly and compares greatly with the rotation rate yielded by {\tt DROP}, with an error below a percent. We see that Roberts' equation compare greatly with the numerical reference, except at short (where a divergence is seen) and large radii (with an error of $\sim 3~\%$).
 
\begin{table}
       \centering
       \begin{tabular}{lrrr}
              &\multicolumn{3}{c}{configuration A} \\\hline    
              & \citet{hachisu86} & {\tt DROP}$^\dagger$ & this work \\
              $\bare_{\rm s}$ & $\leftarrow 0.750$ & $\leftarrow 0.750$ & $\leftarrow 0.750$ \\
              $n$ & $\leftarrow 1.5$ & $\leftarrow 1.5$ & $\leftarrow 1.5$ \\
              $M/\big[\rho_{\rm c}R_{\rm e}^3\big]$ & $0.430$ & $0.43027$ & $0.43280$ \\
              $V/R_{\rm e}^3$ & $3.03$ & $3.02976$  & $3.14159$\\
              $\hatomega^2\times2\uppi$ & $0.227$ & $0.22663$ & $^{\ast}0.22760$ \\
              $J/\big[G\rho_{\rm c}^3R_{\rm e}^{10}\big]^{1/2}$ & $0.0356$ & $0.03556$ & $0.03609$ \\
              $-W/\big[G\rho_{\rm c}^{2}R_{\rm e}^{5}\big]$ & $0.183$ & $0.18345$ & $0.18496$ \\
              $T/\big[G\rho_{\rm c}^{2}R_{\rm e}^{5}\big]$ & $0.00847$ & $0.00846$ & $0.00861$ \\
              $U/\big[G\rho_{\rm c}^{2}R_{\rm e}^{5}\big]$ & $0.167$ & $0.16652$ & $0.16697$\\
              $|{\rm VP}/W|$ & $<10^{-3}$ & $3\cdot10^{-8}$ & $8\cdot10^{-4}$\\
              \hline
              \multicolumn{4}{l}{$\leftarrow$ input data}\\
              \multicolumn{4}{l}{$^\dagger$SCF-method \citep{bh21}}\\
              \multicolumn{4}{l}{$^\ast$Averaged, see \eqref{eq:om2_mean}}\\
       \end{tabular}
       \caption{Configuration A and corresponding global quantities. Results from the tables of \citet{hachisu86} are reported in the first column.}
       \label{tab:configA}
\end{table}

We have calculated the main global properties of the polytrope, namely the mass $M$, the volume $V$ and the angular momentum $J$, the gravitational, kinetic and internal energies, $W$, $T$ and $U$ respectively (see the Appendix \ref{sec:intvol}) and compared with the tables of \citet{hachisu86}. The results are reported in Tab. \ref{tab:configA}. We see that the values obtained are slightly overestimated with the present approximation. This is due to the boundary of the fluid, which is below the corresponding spheroidal surface, as seen from Fig. \ref{fig:configA}c. Thus, the volume of the fluid, and all volume integrals following, are clearly greater than the outputs of the numerical reference. Furthermore, the value of the Virial parameter, i.e. $|{\rm VP}/W|\approx 8\cdot10^{-4}\ll 1$, also validates the approximation in this case.

\subsection{On critical rotations}

We can go further in the comparison by looking at an extreme configuration, i.e. a configuration near the so-called ``critical-rotations'' \citep{hachisu86}, where matter at the surface is bearly bounded to the system. Such objects deviate largely from spheroids and we expect the approximation to fail at this point. 

\begin{table}
       \centering
       \begin{tabular}{lrrr}
              &\multicolumn{3}{c}{configuration B} \\\hline    
              & \citet{hachisu86} & {\tt DROP}$^\dagger$ & this work \\
              $\bare_{\rm s}$ & $\leftarrow 0.662$ & $\leftarrow 0.662$ & $\leftarrow 0.662$ \\
              $n$ & $\leftarrow 3.0$ & $\leftarrow 3.0$ & $\leftarrow 3.0$ \\
              $M/\big[\rho_{\rm c}R_{\rm e}^3\big]$ & $0.0255$ & $0.02545$ & $0.02546$ \\
              $V/R_{\rm e}^3$ & $2.30$ & $2.24640$  & $2.77298$\\
              $\hatomega^2\times2\uppi$ & $0.0256$ & $0.02563$ & $^{\ast}0.02567$ \\
              $J/\big[G\rho_{\rm c}^3R_{\rm e}^{10}\big]^{1/2}$ & $0.00015$ & $0.00016$ & $0.00016$ \\
              $-W/\big[G\rho_{\rm c}^{2}R_{\rm e}^{5}\big]$ & $0.00140$ & $0.00139$ & $0.00139$ \\
              $T/\big[G\rho_{\rm c}^{2}R_{\rm e}^{5}\big]$ & $0.00001$ & $0.00001$ & $0.00001$ \\
              $U/\big[G\rho_{\rm c}^{2}R_{\rm e}^{5}\big]$ & $0.00137$ & $0.00137$ & $0.00137$\\
              $|{\rm VP}/W|$ & $<10^{-3}$ & $1\cdot10^{-8}$ & $8\cdot10^{-7}$\\
              \hline
       \end{tabular}
       \caption{Same legend as Tab. \ref{tab:configA}, but for configuration B.}
       \label{tab:configB}
\end{table}

\begin{figure}
       \centering
       \includegraphics[trim={1mm 4.1cm 10mm 0},width=\linewidth]{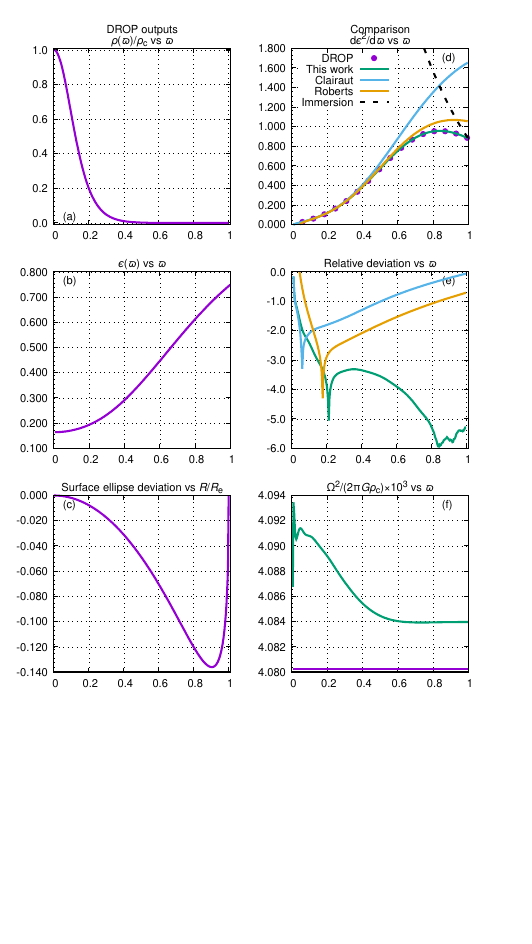}
       \caption{Same legend as for Fig. \ref{fig:configA}, but for configuration B ($\bare_{\rm s}=0.662$, $n=3$). The dashed black line represents the condition for immersion, i.e. \eqref{eq:immersion_e2}.}
       \label{fig:configB}
\end{figure}

We first consider configuration B, with a ``soft'' EoS ($n=3$). The configuration and its global properties are reported in Tab. \ref{tab:configB} and the results are plotted in Fig. \ref{fig:configB}. Surprisingly, the agreement between the spheroidal approximation and the numerical reference is very good; see Fig. \ref{fig:configB}d and Tab. \ref{tab:configB}. For $\varpi>0.2$, we see that the discrepancy is $\lesssim10^{-3}$ in relative. For shorter radii, the gap is wider, due to the numerical precision of the derivatives, as the values themselves are ``small'' ; so any discrepancy is amplified. The approximation seems to stay valid at the surface, even though the deviation from a spheroid is large (see panel c). This can be explained by the mass density curve, namely panel a. Indeed, we see that, for $\varpi>0.4$, we have $\rho(\varpi)\ll\rho_{\rm c}$, so the contribution of this part to the gravitational potential (and thus, to the rotation rate and \eqref{eq:id_nsfoe}) is negligible. So, as long as the isopycnics for $\varpi<0.4$ are close enough to spheroids, the approximation is still valid.

We also have plotted in panel d of Fig. \ref{fig:configB} the upper limit of the immersion criterion, i.e. \eqref{eq:immersion_e2}. Interestingly, the squared eccentricity gradient seems to tend to this limit for $\varpi=1$, i.e. at the surface. This would imply that at the critical rotation, we have $(\du b/\du a)|_{a=R_{\rm e}}\rightarrow0$, where we used the physical radii, namely the matter at the pole is crushed. 

\begin{table}
       \centering
       \begin{tabular}{lrrr}
              &\multicolumn{3}{c}{configuration C} \\\hline    
              & \citet{hachisu86} & {\tt DROP}$^\dagger$ & this work \\
              $\bare_{\rm s}$ & $\leftarrow 0.442$ & $\leftarrow 0.442$ & $\leftarrow 0.442$ \\
              $n$ & $\leftarrow 0.5$ & $\leftarrow 0.5$ & $\leftarrow 0.5$ \\
              $M/\big[\rho_{\rm c}R_{\rm e}^3\big]$ & $0.767$ & $0.77072$ & $0.82697$ \\
              $V/R_{\rm e}^3$ & $1.59$ & $1.56633$  & $1.85144$\\
              $2\uppi\hatomega^2$ & $0.939$ & $0.94158$ & $^{\ast}0.93380$ \\
              $J/\big[G\rho_{\rm c}^3R_{\rm e}^{10}\big]^{1/2}$ & $0.199$ & $0.19936$ & $0.22563$ \\
              $-W/\big[G\rho_{\rm c}^{2}R_{\rm e}^{5}\big]$ & $0.531$ & $0.53568$ & $0.59836$ \\
              $T/\big[G\rho_{\rm c}^{2}R_{\rm e}^{5}\big]$ & $0.0962$ & $0.09672$ & $0.10902$ \\
              $U/\big[G\rho_{\rm c}^{2}R_{\rm e}^{5}\big]$ & $0.339$ & $0.34230$ & $0.35218$\\
              $|{\rm VP}/W|$ & $<10^{-3}$ & $1\cdot10^{-4}$ & $5\cdot10^{-2}$\\
              \hline
       \end{tabular}
       \caption{Same legend as Tab. \ref{tab:configA}, but for configuration C.}
       \label{tab:configC}
\end{table}

\begin{figure}
       \centering
       \includegraphics[trim={1mm 4.1cm 10mm 0},width=\linewidth]{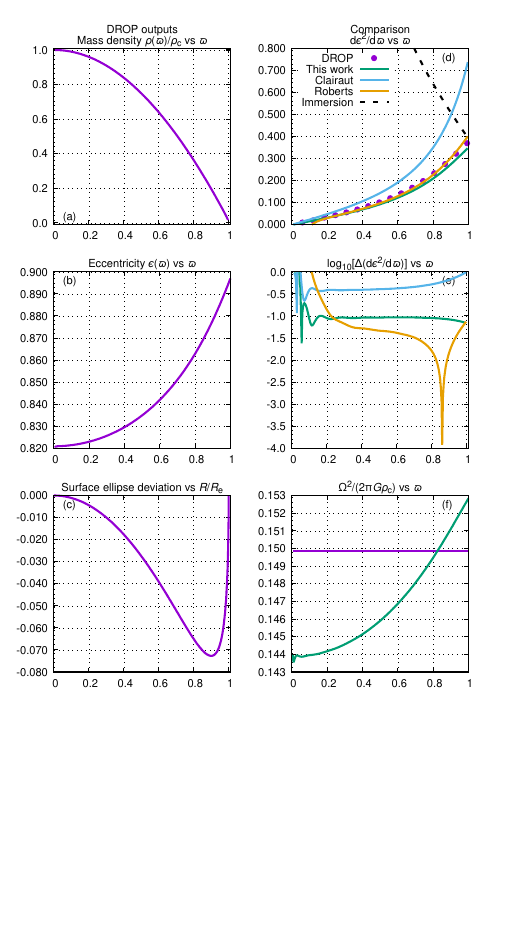}
       \caption{Same legend as for Fig. \ref{fig:configB}, but for configuration C ($\bare_{\rm s}=0.442$, $n=0.5$).}
       \label{fig:configC}
\end{figure}

Another example of critical rotation is displayed in configuration C, where the EoS is ``hard'' ($n=0.5$). The configuration and its global properties are reported in Tab. \ref{tab:configC} and the results are plotted in Fig. \ref{fig:configC}. Here, the agreement between the spheroidal approximation reported here and the numerical reference is not good at all, with a relative error of at least 10\% on $\du\epsilon^2/\du\varpi$ and the global properties. Only the averaged rotation rate is correct, but we see from Fig. \ref{fig:configC}f that the rate itself is not a constant anymore (with an amplitude of, again, $\sim10\%$ of the mean value). This disagreement is explained by the large deviation of the external surface to a spheroid, which is not cancelled by the mass density profile, i.e. $\hatrho(\varpi)\ll1$ only very close to the surface $(\varpi=1)$. So, the deviation from a spheroid has here a real impact, as the gravitational potential arising from this mass distribution is significatively different from the one produced by a spheroidally stratified object. 

However, we observe once again that the immersion criterion joins with the $\du\epsilon^2/\du \varpi$-curve computed from {\tt DROP} at $\varpi=1$, reinforcing our conclusion of the previous example.

\section{The limit of small flattenings}\label{sec:clairaut}

\subsection{The IDE at first order}

The case of slowly rotating structures is of great importance in the context of planetary and stellar interiors \citep[e.g.][]{cr63,zt70}. Such situations suppose that the deviation to sphericity is small, i.e. $\epsilon^2(\varpi)\ll1$. While the Earth or the Sun can probably be considered as slow rotators, this does not seem to be the case of Jupiter and Saturn. The functions defined by \eqref{eq:chi_def}, \eqref{eq:mu_def} and \eqref{eq:nu_def} can then be expanded at first order in $\epsilon^2$. So, we obtain
\begin{align}
       \chi(\varpi',\varpi) &= \frac{\varpi'^3}{2\varpi^6}\left[\varpi'^2\epsilon^2(\varpi')-\varpi^2\epsilon^2(\varpi)\right] + {\cal O}(\epsilon^4),
\end{align}
\begin{align}
  \mu(\varpi',\varpi) &= -\frac13\frac{\varpi'^3}{\varpi^3}\\
  & \times \left[1-\epsilon^2(\varpi')\left(\frac12+\frac35\frac{\varpi'^2}{\varpi^2}\right)+\frac32\epsilon^2(\varpi)\right] + {\cal O}(\epsilon^4),\notag
\end{align}
and
\begin{align}
       \nu(\varpi') &= -\frac13-\frac2{15}\epsilon^2(\varpi')+ {\cal O}(\epsilon^4),
\end{align}
respectively. Thus,  at first order in $\epsilon^2$, \eqref{eq:id_nsfoe} becomes
\begin{align}
  \label{eq:id_e2}
      -\frac13\frac{\du\epsilon^2}{\du \varpi} &\left(\int_{\hatrho(0)}^{\hatrho(\varpi)} \du\hatrho(\varpi') \frac{\varpi'^3}{\varpi^3} + \int_{\hatrho(\varpi)}^{\hatrho(1)} \du\hatrho(\varpi')\right)\\
      &\mkern-5mu\approx\frac1{\varpi^6}\int_{\hatrho(0)}^{\hatrho(\varpi)}\du\hatrho(\varpi') \varpi'^3 \left[\varpi'^2\epsilon^2(\varpi')-\varpi^2\epsilon^2(\varpi)\right]. \notag     
\end{align}
Note that $\du\epsilon^2/\du\varpi$ is already first order in $\epsilon^2$, so the first order terms arising from $\mu$ and $\nu$ can be neglected.

\subsection{Clairaut's equation recovered}

Except in some particular cases, the mass density vanishes continuously at the surface. By integrating (\ref{eq:id_e2}) by parts, we obtain
\begin{align}\label{eq:id_clrt}
  \frac{\varpi^6}{3}\frac{\du\epsilon^2}{\du \varpi}\hatrho_{\rm m} & \approx \varpi^5\epsilon^2(\varpi)\hatrho_{\rm m}\\
  &- \int_0^\varpi\du \varpi'\hatrho(\varpi')\left[5\varpi'^4\epsilon^2(\varpi')+\varpi'^5\frac{\du\epsilon^2}{\du \varpi'}\right] \notag,
\end{align}
where 
\begin{equation}
       \hatrho_{\rm m}(\varpi) = \frac{3}{\varpi^3}\int_0^\varpi \du \varpi' \hatrho(\varpi')\varpi'^2,
\end{equation}
is the classicaly called the mean density \citep[e.g.][]{tisserand91,ragazzo2020}, evaluated from the center to the running radius. In this form, \eqref{eq:id_clrt} is suitable to eliminate the integral by differentiation. So, we derivate a second time with respect to the physical radius $a=R_{\rm e}\varpi$ to obtain
\begin{equation}\label{eq:clrt_e2}
  \frac{\du^2\epsilon^2}{\du a^2} + \frac6a\frac{\hatrho}{\hatrho_{\rm m}}\frac{\du\epsilon^2}{\du a} + \frac{6}{a^2}\left(\frac{\hatrho}{\hatrho_{\rm m}}-1\right)\epsilon^2 \approx 0. 
\end{equation}
This result clearly recalls the fundamental equation derived by \citet{clairaut43}, namely
\begin{equation}\label{eq:clrt}
  \frac{\du^2f}{\du b^2} + \frac6b\frac{\hatrho}{\langle\hatrho\rangle}\frac{\du f}{\du b} + \frac{6}{b^2}\left(\frac{\hatrho}{\langle\hatrho\rangle}-1\right) f = 0. 
\end{equation}
where $f= 1-\sqrt{1-\epsilon^2}$ is the flattening of the isopycnic surface, $b$ is its polar radius and 
\begin{equation}
  \langle\hatrho\rangle = \frac{3}{b^3}\int_0^b \du b' \hatrho(b')b'^2.
\end{equation}
Let us show that \eqref{eq:clrt_e2} and \eqref{eq:clrt} are fully compatible. At first order in $\epsilon^2$, we have ${2f\approx \epsilon^2}$ and ${b \approx a(1-\epsilon^2/2)}$, so
\begin{equation}
  2\frac{\du f}{\du b} \approx \frac{\du\epsilon^2}{\du a}\left[1+\frac12\left(\epsilon^2+a\frac{\du\epsilon^2}{\du a}\right)\right] \approx \frac{\du\epsilon^2}{\du a}.
\end{equation}
Now, as the derivatives and the function $f$ itself are already of first order in $\epsilon^2$, only the ``zeroth'' order in $\langle\hatrho\rangle$ is needed. At this order, we have $a\approx b$ and thus $\langle\hatrho\rangle\approx\hatrho_{\rm m}$. Hence, we conclude that \eqref{eq:id_nsfoe} is equivalent to Clairaut's differential equation in the limit of small flattenings, at first order in $\epsilon^2$. Note that some authors \citep[e.g.][]{ragazzo2020} use the mean radius $(a^2b)^{1/3}$ instead of $a$ or $b$. We can show by the same reasoning that the equations would still agree at first order.

\subsection{An example. Comparison with Chandrasekhar's pertubative approach}

\begin{figure}
       \centering
       \includegraphics[trim={1mm 4.1cm 10mm 0},clip,width=\linewidth]{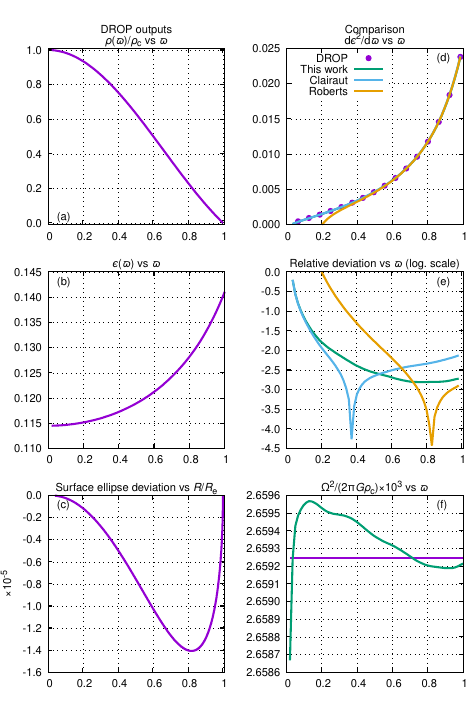}
       \caption{Same legend as Fig. \ref{fig:configA}, but for configuration D (${\bare_{\rm s}=0.99}$, ${n=1.0}$).}
       \label{fig:configD}
\end{figure}

To illustrate the compatibility between Clairaut's equation and \eqref{eq:id_nsfoe}, let us consider the numerical solution computed from {\tt DROP} for a self-gravitating polytrope with $\bare_{\rm s}=0.99$ and $n=1$, hereafter configuration D ; see Tab. \ref{tab:configD} for the details of the configuration and the associated global quantities. We have $\epsilon_{\rm s}^2=0.0199$, which is expected to be ``small enough'' for the expansions made in the previous paragraph to be valid. We can therefore check our expansions as well as Clairaut's equation. The results are presented in Fig. \ref{fig:configD} (same panels as for configuration A). We notice that the $\epsilon$-profile is close to a quadratic. We see, again, the excellent agreement between the present approach and Clairaut's equation. Also, we see that the global quantities obtained with the IDE are close to the one obtained with {\tt DROP}, with between four to six digits shared on the values. As quoted in the introduction, this is not a surprise, as \citet{clairaut43} showed that \emph{for small deviations from the sphere}, i.e. small flattenings, \emph{the isopycnic surfaces are ellipses in any meridian plane}. 

Moreover, slowly rotating polytropes have been studied by many authors, in particular by \citet{ch33}. His approach is based on the Lane-Emden equation, supplemented by a small amplitude, rotational field. The equilibrium is solved in the form of series. Configurations with $n=1$ (like configuration B) are interesting because the results arising from this theory are purely analytical and offer a interesting opportunity for comparisons. As the dimensionless rotation rate $\hatomega^2$ is an input in Chandrasekhar's work (while the axis ratio $\bare$ is an output), the comparison is performed by injecting the rotation rate provided by {\tt DROP} into Chandrasekhar's equations. The results are reported in Tab. \ref{tab:configD} (column 2). We see that the comparison is satisfactory, the agreement being much better than $1 \%$. Furthermore, the Virial quantities yielded by the spheroidal approximation are in excellent agreement with the numerical reference.

\begin{table}
       \centering
       \begin{tabular}{lrrr}
              &\multicolumn{3}{c}{configuration D} \\\hline    
              & \citetalias{ch33}& {\tt DROP}$^\dagger$ & this work \\
              $\bare_{\rm s}$ & $0.99008$ & $\leftarrow 0.990$ & $\leftarrow 0.990$ \\
              $n$ & $\leftarrow 1.0$ & $\leftarrow 1.0$ & $\leftarrow 1.0$ \\
              $M/\big[\rho_{\rm c}R_{\rm e}^3\big]$ & $1.25799$ & $1.25807$ & $1.25806$ \\
              $V/R_{\rm e}^3$ & $4.14641$ & $4.14683$  & $4.14690$\\
              $\hatomega^2\times2\uppi$ & $\leftarrow0.01671$ & $0.01671$ & $^{\ast}0.01671$ \\
              $J/\big[G\rho_{\rm c}^3R_{\rm e}^{10}\big]^{1/2}$ & $ $ & $0.04244$ & $0.04244$ \\
              $-W/\big[G\rho_{\rm c}^{2}R_{\rm e}^{5}\big]$ & $ $ & $1.19148$ & $1.19148$ \\
              $T/\big[G\rho_{\rm c}^{2}R_{\rm e}^{5}\big]$ & $ $ & $0.00274$ & $0.00274$ \\
              $U/\big[G\rho_{\rm c}^{2}R_{\rm e}^{5}\big]$ & $ $ & $1.18600$ & $1.18599$\\
              $|{\rm VP}/W|$ & $ $ & $7\cdot10^{-12}$ & $1\cdot10^{-5}$\\
              \hline
       \end{tabular}
       \caption{Same legend as for Tab. \ref{tab:configA} but for configuration D, which is compared with \citet{ch33} (first column).}
       \label{tab:configD}
\end{table}

\section{Introduction of mass-density jumps: the modified IDE}
\label{sec:jumps}

Mass-density jumps are usually associated with a sudden change in the equation of state or in the mechanism transporting matter or energy. It is therefore interesting to render the present method as flexible as possible, and to account for such discontinuities. As often, we consider jumps as zero-thickness transitions, while, in real systems, these have always have certain spatial extension. Inspired by Sec. \ref{ssec:maclaurin}, we can easily introduce mass-density jumps in the present formalism by decomposing the mass-density profile as
\begin{equation}\label{eq:rho_jumps}
       \hat\rho(\varpi) = \sum_{k=1}^{\cal K} \left[\hat\rho_k(\varpi)-\hat\rho_{k+1}(\varpi)\right]{\cal H}(\varpi_k-\varpi),
\end{equation}
where $\cal K$ is the number of \emph{heterogeneous domains} and $\hat\rho_k(\varpi)=\rho_k(\varpi)/\rho_c$ is the mass density inside domain number $k$ (we still normalise mass-densities to the central value $\rho_c$). As for the discrete case, we have set ${\hat\rho_{{\cal K}+1}(\varpi)=0}$ to keep a single sum, which means that the outer space is the very last domain, with index ${\cal K}+1$ and null mass density. There are therefore ${\cal K}$ jumps, located at ${\varpi=\varpi_k,} \; k\in\llbra 1,{\cal K} \rrbra$. Note that (\ref{eq:rho_jumps}) allows for configurations with a surface discontinuity, i.e. at ${\varpi_{\cal K}=1}$. The derivative of this profile writes 
\begin{align}\label{eq:drho_jumps}
       \frac{\du\hat\rho}{\du \varpi} = \sum_{k=1}^{\cal K} \bigg[\frac{\du\hat\rho_k}{\du \varpi}&-\frac{\du\hat\rho_{k+1}}{\du \varpi}\bigg]{\cal H}(\varpi_k-\varpi) \notag\\ &- \sum_{k=1}^{\cal K} \left[\hat\rho_k(\varpi)-\hat\rho_{k+1}(\varpi)\right]\delta(\varpi_k-\varpi).
\end{align}
We can thus make use of the properties of the Heaviside and Dirac distributions to generalise \eqref{eq:id_nsfoe}.

\subsection{Piece-wise rotation and discontinuity in the ellipticity}

Let us consider that each domain $k \in \llbra 1,\cal K\rrbra$ rotates rigidly at its own rate $\hatomega_k$. So, for a given domain $k_0$, we have ${\varpi\in{]\varpi_{k_0-1},\varpi_{k_0}[}}$, and \eqref{eq:drhoom2_nsfoe} becomes\footnote{We have introduced $\varpi_0=0$ for convenience.}
\begin{align}
  \label{eq:drhoom2_jumps2}
  -\hatomega_{k_0}^2&= \sum_{k=1}^{k_0-1}\int_{\varpi_{k-1}}^{\varpi_k}\du\varpi'\frac{\du\hatrho_k}{\du\varpi'} \kappa^{\rm in}(\varpi',\varpi) \\
       &+ \int_{\varpi_{k_0-1}}^{\varpi} \du\varpi'\frac{\du\hatrho_k}{\du\varpi'} \kappa^{\rm in}(\varpi',\varpi)\notag\\
  &+ \int_\varpi^{\varpi_{k_0}} \du\varpi'\frac{\du\hatrho_k}{\du\varpi'} \kappa^{\rm out}(\varpi',\varpi)\notag\\
  & + \sum_{k=k_0}^{\cal K}\int_{\varpi_{k-1}}^{\varpi_k}\du\varpi'\frac{\du\hatrho_k}{\du\varpi'} \kappa^{\rm out}(\varpi',\varpi)\notag\\
  &-\sum_{k=1}^{k_0-1}\frac{\alpha_k-1}{\alpha_k}\hatrho_k(\varpi_k) \kappa^{\rm in}(\varpi_k,\varpi)\notag\\
       &-\sum_{k=k_0}^{\cal K}\frac{\alpha_k-1}{\alpha_k}\hatrho_k(\varpi_k)\kappa^{\rm out}(\varpi_k,\varpi),\notag
\end{align}
where $\alpha_k = \hatrho_k(\varpi_k)/\hatrho_{k+1}(\varpi_k)$ is the mass density jump at each interface $k$. A major question concerns the behavior of this equation when applied to two adjacent domains. To answer this point, we write \eqref{eq:drhoom2_jumps2} at $\varpi_-=\varpi_{k_0}-\Delta\varpi$ (inside layer $k_0$) and at $\varpi_+=\varpi_{k_0}+\Delta\varpi$ (inside layer $k_0+1$), with $\Delta\varpi>0$.
In the limit where $\Delta\varpi\ll1$, the difference in the rotation rates between $\varpi_-$ and $\varpi_+$ satisfies
\begin{align}
       &\hatomega_{k_0}^2-\hatomega_{k_0+1}^2 = 4\ \Delta\varpi \times \label{eq:absurd}\\
       &\quad\Bigg\{\sum_{k=1}^{k_0}\int_{\varpi_{k-1}}^{\varpi_k}\du\varpi'\frac{\du\hatrho_k}{\du\varpi'}\left[2\chi(\varpi',\varpi_{k_0})-\frac{\du\epsilon^2}{\du\varpi}\Bigg|_{\varpi_{k_0}}\mu(\varpi',\varpi_{k_0})\right]\notag\\
       &\mkern+150mu-\sum_{k=k_0+1}^{\cal K}\int_{\varpi_{k-1}}^{\varpi_k}\du\varpi'\frac{\du\hatrho_k}{\du\varpi'}\frac{\du\epsilon^2}{\du\varpi}\Bigg|_{\varpi_{k_0}}\nu(\varpi')\notag\\
       &\quad-\sum_{k=1}^{k_0}\frac{\alpha_k-1}{\alpha_k}\hatrho_k(\varpi_k)\left[2\chi(\varpi_k,\varpi_{k_0})-\frac{\du\epsilon^2}{\du\varpi}\Bigg|_{\varpi_{k_0}}\mu(\varpi_k,\varpi_{k_0})\right] \notag\\
       &\mkern+150mu+\sum_{k=k_0+1}^{\cal K}\frac{\alpha_k-1}{\alpha_k}\hatrho_k(\varpi_k)\frac{\du\epsilon^2}{\du\varpi}\Bigg|_{\varpi_{k_0}}\nu(\varpi_k)\Bigg\}\notag,
\end{align}
at first order in $\Delta\varpi$. If asynchroneous motion is possible, then the RHS of this expression must remain finite when $\Delta\varpi \rightarrow 0$. We see from (\ref{eq:absurd}) that this is possible {\it only if the eccentricity undergoes a discontinuity} at $\varpi_{k_0}$, namely
\begin{equation}
       \frac{\du\epsilon^2}{\du\varpi}\Bigg|_{\varpi_{k_0}} = \frac{\epsilon^2_{k_0+1}(\varpi_{k_0})-\epsilon^2_{k_0}(\varpi_{k_0})}{\Delta\varpi},
\end{equation}
where $\epsilon_{k_0}$ is the eccentricity profile in the domain $k_0$. Note that \eqref{eq:absurd} can not be used to quantify this jump, as we assumed a continuous eccentricity to arrive at this point. Indeed, if these jumps are considered from the beginning, they would cause discontinuities in the $\kappa$-functions, $\chi$, $\mu$ and $\nu$, which makes the calculations far more complex.

This ``eccentricity jump'' only states that \emph{the interfaces between layers are not isopycnic surfaces}. The isopycnic in the inner layer (the ``core'') intersect the interface and is prolongated by another isopycnic in the outer layer (the ``envelope'') whose eccentricity has no reason to be the same. This statement has two interesting consequences: {i)} the ``eccentricity jump'' occurs not at a single value of $\varpi$ but on a whole range close to any interface ; {ii)} the potential of an incomplete Maclaurin spheroid being unknown analytically, \emph{the continuous version of the NSFoE cannot describe systems with rotational discontinuities}.

\subsection{Global, rigid rotation}

By requiring $\hatomega_{k}=\hatomega, \ \forall k\in\llbra1,{\cal K}\rrbra$, both sides in \eqref{eq:absurd} are null in the limit $\Delta \varpi\rightarrow0$, meaning \emph{no eccentricity jump occurs for systems in global rotation}, so that \emph{the interfaces between layers are isopycnic surfaces}. As the RHS of \eqref{eq:drhoom2_jumps2} is constant, we can, as in the single-layer case, take its derivative with respect to $\varpi$ inside layer $k_0$. We find
\begin{align}
  \label{eq:id_jumps}
  &\sum_{k=1}^{k_0-1}\int_{\varpi_{k-1}}^{\varpi_k}\du\varpi'\frac{\du\hatrho_k}{\du\varpi'} \chi(\varpi',\varpi) + \int_{\varpi_{k_0-1}}^{\varpi}\du\varpi'\frac{\du\hatrho_k}{\du\varpi'}\chi(\varpi',\varpi)\notag\\
       &\mkern+50mu-\sum_{k=1}^{k_0-1}\frac{\alpha_k-1}{\alpha_k}\hatrho_k(\varpi_k)\chi(\varpi_k,\varpi)=\frac12\frac{\du\epsilon^2}{\du \varpi}\times \\
       &\Bigg\{\sum_{k=1}^{k_0-1}\int_{\varpi_{k-1}}^{\varpi_k} \du\varpi'\frac{\du\hatrho_k}{\du\varpi'}\mu(\varpi',\varpi) + \int_{\varpi_{k_0-1}}^{\varpi} \du\varpi'\frac{\du\hatrho_k}{\du\varpi'}\mu(\varpi',\varpi)\notag\\
       &\mkern+5mu-\sum_{k=1}^{k_0-1}\frac{\alpha_k-1}{\alpha_k}\hatrho_k(\varpi_k)\mu(\varpi_k,\varpi)- \sum_{k=k_0}^{\cal K}\frac{\alpha_k-1}{\alpha_k}\hatrho_k(\varpi_k)\nu(\varpi_k)\notag\\
       &\mkern+10mu+\sum_{k=k_0+1}^{\cal K}\int_{\varpi_{k-1}}^{\varpi_k} \du\varpi'\frac{\du\hatrho_k}{\du\varpi'}\nu(\varpi')+ \int_{\varpi}^{\varpi_{k_0}} \du\varpi'\frac{\du\hatrho_k}{\du\varpi'}\nu(\varpi')\Bigg\}.\notag
\end{align} 
This expression is the IDE modified by the presence of jumps. Note that it can be recast in the form of (\ref{eq:id_nsfoe_bis}). As for the single-layer case, \eqref{eq:id_jumps} can not be solved alone as we have a single integro-differential equation for ${\cal K}+1$ unknown functions, namely the mass-density profiles $\hatrho_k(\varpi)$ and the eccentricity $\epsilon(\varpi)$. A solution requires $\cal K$ equations of state and $\cal K$ Bernoulli's equations.

\subsection{An example}

Once again, we check the self-consistency of \eqref{eq:id_jumps} by comparison with a numerical solution from {\tt DROP}; see Subsec. \ref{subsect:IDEcheck}. We see that \eqref{eq:id_jumps} can be written in the form of \eqref{eq:id_nsfoe_bis}, i.e. we can obtain an equation of the form $\du\epsilon^2/\du\varpi = g(\rho,\epsilon,\varpi)$. So, as before, we use {\tt DROP} outputs to compute both sides of \eqref{eq:id_jumps} and we then compare the results. 

\begin{figure}
       \centering
       \includegraphics[width=\linewidth]{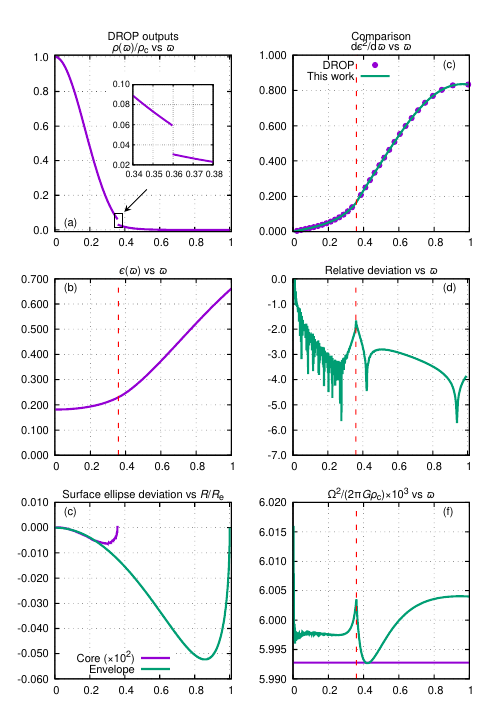}
       \caption{Same legend as for Fig. \ref{fig:configA}, but for configuration A', which is a two-domain body (i.e. ${\cal K}=2$) with a mass-density jump at $\varpi=\varpi_1\approx 0.36$ (marked with a vertical red dshed line; see Tab. \ref{tab:configAp}).}
       \label{fig:configAp}
\end{figure}

\begin{table}
       \centering
       \begin{tabular}{lrr}
              &\multicolumn{2}{c}{configuration A'} \\\hline    
              & {\tt DROP}$^\dagger$ & this work \\
              $\bare_{\rm s}$ & \multicolumn{2}{c}{$\leftarrow 0.750$} \\
              $n_1$ & \multicolumn{2}{c}{$\leftarrow 1.5$}\\
              $n_2$ & \multicolumn{2}{c}{$\leftarrow 3.0$}\\
              $\varpi_1\bare(\varpi_1)$ & \multicolumn{2}{c}{$\leftarrow 0.35$}\\
              $\varpi_1$ & \multicolumn{2}{c}{$0.35965$}\\
              $\alpha_1$ & \multicolumn{2}{c}{$\leftarrow 2.0$}\\
              $M/\big[\rho_{\rm c}R_{\rm e}^3\big]$ & $0.05722$ & $0.05735$ \\
              $V/R_{\rm e}^3$ & $2.92254$  & $3.14519$\\
              $2\uppi\hatomega^2$ & $0.03765$ & $^{\ast}0.03768$ \\
              $J/\big[G\rho_{\rm c}^3R_{\rm e}^{10}\big]^{1/2}$ & $0.00048$ & $0.00048$ \\
              $-W/\big[G\rho_{\rm c}^{2}R_{\rm e}^{5}\big]$ & $0.00631$ & $0.00633$ \\
              $T/\big[G\rho_{\rm c}^{2}R_{\rm e}^{5}\big]$ & $0.00004$ & $0.00004$ \\
              $U/\big[G\rho_{\rm c}^{2}R_{\rm e}^{5}\big]$ & $0.00622$ & $0.00623$\\
              $|{\rm VP}/W|$ & $8\cdot10^{-5}$ & $1\cdot10^{-3}$\\
              \hline
       \end{tabular}
       \caption{Same legend as for Tab. \ref{tab:configB} but for configuration A'.}
       \label{tab:configAp}
\end{table}

 Configuration A' is a rotating body with surface axis ratio of $0.75$, a core with polytropic index $n_1=1.5$ and semi-polar axis $\varpi_1\bare(\varpi_1)=0.35$ and an envelope with polytropic index $n_2=3$. This system could correspond to a highly flatten object with a large convective core (whose mean radius is $\sim40\%$ of the star's radius) and a big radiative envelope; it may thus be considered as a very simple model for a fast-rotating high-mass star ($M\gtrsim1.2~{\rm M_{\odot}}$); see e.g. \citet{maeder2009}. The global quantities are given in Tab \ref{tab:configAp} and the results are plotted in Fig. \ref{fig:configAp}. Again, the agreement between the spheroidal approximation reported here and the numerical reference is remarkable, within a few tenths of a percent (except for the volume). The relative Virial parameter is also really good ($10^{-3}\ll 1$), which validates more the approach. We see that both squared excentricity gradients compare really well to each other (see panel d), the discrepancy being around $\sim 10^{-3}$ in most of the object and around $\sim 10^{-2}$ in the neighboring of the mass density jump, which is due to the numerical resolution in this region. Indeed, for each cylindrical radius, the interface is described by two or three points, which may not be enough to reach a good accuracy on the dynamics of the eccentricity in this region. This peak is also seen in the $\hatomega^2$ curve (panel f), where the gap to the value yielded by {\tt DROP} is also about a few tenth of a percent.

\section{Discussion}
\label{sec:conclusion}

\subsection{Summary}

This article inverstigates the condition of equilibrium of a heterogeneous system with spheroidal isopycnic surfaces (axisymmetrical case). We have derived the main integro-differential equation (IDE) of the problem in the case where the rotation rate is constant onto the isopycnic surfaces, and we have deduced the corresponding IDE in the special case of rigid rotation. This IDE works for a wide range of rotation rates, not only in the slow rotating limit as often considered. Using the {\tt DROP}-code as a numerical reference, we have proven the reliability of the approach for various configurations, including fast rotators; see configuration A and B. The IDE is fully compatible with Clauraut'equation in the case of slow rotation. Furthermore, we have seen a correlation between the state of critical rotation and the criterion of non-intersection of the isopycnics. As shown, mass-density jumps can be taken into account in the model as long as there are no rotational discontinuities.

\subsection{Open questions and perspectives}

\begin{enumerate}

\item[1.] {\bf Rotational discontinuities.} When rotational discontinuities are present, an eccentricity jump is mandatory, meaning the interfaces between layers were not isopycnic surfaces. The approximation of spheroidal isopycnic then fails in this case. However, if the rotational discontinuities (or equivalently, the eccentricity jumps) are small enough, it should be possible to derive an IDE for this case, as the range where the jump occurs becomes negligible. This point would merit an additional work. \\

\item[2.] {\bf From slow to fast rotator: a criterion.} In the limit of small flattenings, our approach compares really well with the one developped by \citet{ch33} and we were able to recover Clairaut's equation at first order in $\epsilon^2$. This adresses the question of the limit between slow rotators (well described by Clairaut's theory) and fast rotators, which can be roughly answered as follows. Let us develop $\nu$ at second order in $\epsilon^2$ (for convienience, we use $\nu$ instead of $\chi$ or $\mu$ as it is a function of a single variable). From \eqref{eq:nu_def}, we directly obtain
\begin{equation}
       \nu(\varpi) = -\frac13-\frac{2}{15}\epsilon^2(\varpi)-\frac{8}{105}\epsilon^4(\varpi) + {\cal O}(\epsilon^6).
\end{equation}

Now, let $\eta$ be the ratio of the fourth order term to the second order term. We have
\begin{equation}
       \eta = \frac47\epsilon^2(\varpi).
\end{equation}

So, roughly, the error in the quantity $\du\epsilon^2/\du\varpi$ made by using Clairaut's equation, i.e. \eqref{eq:clrt_e2}, is of the order of $\eta$. The corresponding axis ratio at the surface is then 
\begin{equation}\label{eq:crit_es}
       \bare_{\rm s} \geq \sqrt{1-\frac74\eta}.
\end{equation}

We see that for configuration D (Fig. \ref{fig:configD}d), which has an axis ratio of $0.99$ at the surface, the maximum error is of order $10^{-2}$ (we do not take into account the part ${\varpi<0.2}$ which is dominated by the errors of the finite-difference scheme). This then corresponds to the criterion \eqref{eq:crit_es}.\\

\item[3.] {\bf Can we expand the IDE at higher orders ?} As shown, expanding the IDE at first-order in $\epsilon^2$ leads to Clairaut's equation. It would then be interesting to derive a second-order Clairaut equation, basically by exanding the IDE a second-order expansion in $\epsilon^2$. This would be another approach to the expansion of Clairaut's equation than \citet{lan62,lan74} who has performed a multipolar expansion of the shape of the object. However, preliminary calculations indicate that the problem might not be any easier than the equation set reported here. This point is still under investigation.\\

\item[4.] {\bf Do exact solutions to the IDE exist ?} As it is well known, analytical solutions are always powerful tools for making models and diagnosis tools, regarding observations. The existence of analytical solutions to the IDE in the form $\rho(\epsilon)$ would be very interesting, and it already represents an exciting perspective. Clairaut's equation is known to have a few analytical solution \citep[e.g.][]{tisserand91,mar00}. Given the complexity of the IDE, we expect any analytical solution to be only approximate. Solutions via a series expansion or linearisation for example would be interesting to seek for.\\

\item[5.] {\bf Towards 2D-structures ?} As quoted, \eqref{eq:id_nsfoe} is not sufficient in itself to derive models for interiors of rotating bodies; it is the case of Clairaut's equation as well. The IDE has to be combined with an EOS and to Bernoulli's equation. However, the IDE enables to reduce the number of dimensions of the problem, from two to one, through the relationship $\rho(\epsilon)$. The computation of the gravitational potential is skipped in this process (in fact, it is already incorporated in the IDE). This is very attractive, in particular in terms of computing time if a large number of structures have to be computed (see below). We are currently preparing an article dealing with the structure of spheroidal stars and planets from a SCF-method \citep{hachisu86} through this dimension reduction.\\

\item[6.] {\bf Inverse problems.} Planets like Jupiter and Saturn do probably not belong to the category of slow rotators. The IDE could therefore be of great help in generating fast internal 2D-structures (with appropriate EOS), under the conditions of the hypothesis of the NSFoE. Next, it would be easy to compute the gravitational moments and to isolate solutions that match the values ``measured'' by space probes. Yet, as pictured by e.g. \citet{mv23}, high-order gravitational moments mostly describe the outer layers of the object, which is the most poorly described zone by the theory reported here; see also \citet{bh23} (Paper III) and references therein. As such, we expect only the first two moments to be accurate enough. Furthermore, as quoted by \citet{net21}, the Concentric Maclaurin Spheroid (CMS) method by \citep{hub13} has high computational needs, meaning that a scan of a given parameter space is tedious. With a very fast algorithm, it could be possible to identify places in the parameter space compatible with the measured $J_{2n}$, which could be further studied with more sophisticated algorithms (e.g. the CMS-method). Obviously, in the case of gaseous planets, the presence of complex winds at the very surface is not strictily compatible with the NSFoE \citep[the 3D-structure of a gaseous planet with zonal winds has been studied by][]{kzs16}. This is worst in stars where meridional circulations are present \citep[see e.g.][]{zahn92}.

\end{enumerate}

\section*{Data availability} All data are incorporated into the article.

\section*{Acknowledgements} We are grateful to A. Albouy, G. Bou{\'e} and M. Serrero for stimulating discussions during our visit at IMCCE in May 2023. 

\bibliographystyle{mnras}

\begin{thebibliography}{}
       \makeatletter
       \relax
       \def\mn@urlcharsother{\let\do\@makeother \do\$\do\&\do\#\do\^\do\_\do\%\do\~}
       \def\mn@doi{\begingroup\mn@urlcharsother \@ifnextchar [ {\mn@doi@}
         {\mn@doi@[]}}
       \def\mn@doi@[#1]#2{\def\@tempa{#1}\ifx\@tempa\@empty \href
         {http://dx.doi.org/#2} {doi:#2}\else \href {http://dx.doi.org/#2} {#1}\fi
         \endgroup}
       \def\mn@eprint#1#2{\mn@eprint@#1:#2::\@nil}
       \def\mn@eprint@arXiv#1{\href {http://arxiv.org/abs/#1} {{\tt arXiv:#1}}}
       \def\mn@eprint@dblp#1{\href {http://dblp.uni-trier.de/rec/bibtex/#1.xml}
         {dblp:#1}}
       \def\mn@eprint@#1:#2:#3:#4\@nil{\def\@tempa {#1}\def\@tempb {#2}\def\@tempc
         {#3}\ifx \@tempc \@empty \let \@tempc \@tempb \let \@tempb \@tempa \fi \ifx
         \@tempb \@empty \def\@tempb {arXiv}\fi \@ifundefined
         {mn@eprint@\@tempb}{\@tempb:\@tempc}{\expandafter \expandafter \csname
         mn@eprint@\@tempb\endcsname \expandafter{\@tempc}}}
       
       \bibitem[\protect\citeauthoryear{{Basillais} \& {Hur{\'e}}}{{Basillais} \&
         {Hur{\'e}}}{2021}]{bh21}
       {Basillais} B.,  {Hur{\'e}} J.-M.,  2021, \mn@doi [\mnras]
         {10.1093/mnras/stab1658}, \href
         {https://ui.adsabs.harvard.edu/abs/2021MNRAS.506.3773B} {506, 3773}
       
       \bibitem[\protect\citeauthoryear{{Basillais} \& {Hur{\'e}}}{{Basillais} \&
         {Hur{\'e}}}{2023}]{bh23}
       {Basillais} B.,  {Hur{\'e}} J.-M.,  2023, \mn@doi [\mnras]
         {10.1093/mnras/stad151}, \href
         {https://ui.adsabs.harvard.edu/abs/2023MNRAS.520.1504B} {520, 1504}
       
       \bibitem[\protect\citeauthoryear{{Bizyaev}, {Borisov}  \& {Mamaev}}{{Bizyaev}
         et~al.}{2015}]{bbm15}
       {Bizyaev} I.~A.,  {Borisov} A.~V.,   {Mamaev} I.~S.,  2015, \mn@doi [Celestial
         Mechanics and Dynamical Astronomy] {10.1007/s10569-015-9608-5}, \href
         {https://ui.adsabs.harvard.edu/abs/2015CeMDA.122....1B} {122, 1}
       
       \bibitem[\protect\citeauthoryear{{Carciofi}, {Domiciano de Souza},
         {Magalh{\~a}es}, {Bjorkman}  \& {Vakili}}{{Carciofi} et~al.}{2008}]{cado2008}
       {Carciofi} A.~C.,  {Domiciano de Souza} A.,  {Magalh{\~a}es} A.~M.,  {Bjorkman}
         J.~E.,   {Vakili} F.,  2008, \mn@doi [\apjl] {10.1086/586895}, \href
         {https://ui.adsabs.harvard.edu/abs/2008ApJ...676L..41C} {676, L41}
       
       \bibitem[\protect\citeauthoryear{{Chandrasekhar}}{{Chandrasekhar}}{1933}]{ch33}
       {Chandrasekhar} S.,  1933, \mn@doi [\mnras] {10.1093/mnras/93.5.390}, \href
         {https://ui.adsabs.harvard.edu/abs/1933MNRAS..93..390C} {93, 390}
       
       \bibitem[\protect\citeauthoryear{{Chandrasekhar}}{{Chandrasekhar}}{1969}]{chandra69}
       {Chandrasekhar} S.,  1969, {Ellipsoidal figures of equilibrium}.
       Yale Univ. Press, New Haven
       
       \bibitem[\protect\citeauthoryear{{Chandrasekhar} \& {Roberts}}{{Chandrasekhar}
         \& {Roberts}}{1963}]{cr63}
       {Chandrasekhar} S.,  {Roberts} P.~H.,  1963, \mn@doi [\apj] {10.1086/147686},
         \href {https://ui.adsabs.harvard.edu/abs/1963ApJ...138..801C} {138, 801}
       
       \bibitem[\protect\citeauthoryear{Clairaut}{Clairaut}{1743}]{clairaut43}
       Clairaut A.~C.,  1743, Th{\'e}orie de la figure de la Terre tir{\'e}e des
         principes de l'hydrostatique.
       David Fils, Paris
       
       \bibitem[\protect\citeauthoryear{{Domiciano de Souza} et~al.,}{{Domiciano de
         Souza} et~al.}{2014}]{dkm14}
       {Domiciano de Souza} A.,  et~al., 2014, \mn@doi [\aap]
         {10.1051/0004-6361/201424144}, \href
         {https://ui.adsabs.harvard.edu/abs/2014A&A...569A..10D} {569, A10}
       
       \bibitem[\protect\citeauthoryear{{Fujisawa} \& {Eriguchi}}{{Fujisawa} \&
         {Eriguchi}}{2014}]{fe14}
       {Fujisawa} K.,  {Eriguchi} Y.,  2014, \mn@doi [\mnras] {10.1093/mnrasl/slt159},
         \href {https://ui.adsabs.harvard.edu/abs/2014MNRAS.438L..61F} {438, L61}
       
       \bibitem[\protect\citeauthoryear{{Hachisu}}{{Hachisu}}{1986}]{hachisu86}
       {Hachisu} I.,  1986, \mn@doi [\apjs] {10.1086/191121}, \href
         {http://cdsads.u-strasbg.fr/abs/1986ApJS...61..479H} {61, 479}
       
       \bibitem[\protect\citeauthoryear{Hamy}{Hamy}{1890}]{hamy90}
       Hamy M.,  1890, Journal de math\'ematiques pures et appliqu\'ees 4e s{\'e}rie, 6, 69
       
       \bibitem[\protect\citeauthoryear{{Hubbard}}{{Hubbard}}{2013}]{hub13}
       {Hubbard} W.~B.,  2013, \mn@doi [\apj] {10.1088/0004-637X/768/1/43}, \href
         {https://ui.adsabs.harvard.edu/abs/2013ApJ...768...43H} {768, 43}
       
       \bibitem[\protect\citeauthoryear{{Hur{\'e}}}{{Hur{\'e}}}{2022a}]{h2022a}
       {Hur{\'e}} J.-M.,  2022a, \mn@doi [\mnras] {10.1093/mnras/stab3388}, \href
         {https://ui.adsabs.harvard.edu/abs/2022MNRAS.512.4031H} {512, 4031} (Paper I)
       
       \bibitem[\protect\citeauthoryear{{Hur{\'e}}}{{Hur{\'e}}}{2022b}]{h2022b}
       {Hur{\'e}} J.-M.,  2022b, \mn@doi [\mnras] {10.1093/mnras/stac521}, \href
         {https://ui.adsabs.harvard.edu/abs/2022MNRAS.512.4047H} {512, 4047} (Paper II)
       
       \bibitem[\protect\citeauthoryear{{Hur{\'e}} \& {Hersant}}{{Hur{\'e}} \&
         {Hersant}}{2017}]{hh17}
       {Hur{\'e}} J.-M.,  {Hersant} F.,  2017, \mn@doi [\mnras]
         {10.1093/mnras/stw2569}, \href
         {https://ui.adsabs.harvard.edu/abs/2017MNRAS.464.4761H} {464, 4761}
       
       \bibitem[\protect\citeauthoryear{{Kong}, {Zhang}  \& {Schubert}}{{Kong}
         et~al.}{2015}]{kzs15}
       {Kong} D.,  {Zhang} K.,   {Schubert} G.,  2015, \mn@doi [Physics of the Earth
         and Planetary Interiors] {10.1016/j.pepi.2015.09.008}, \href
         {https://ui.adsabs.harvard.edu/abs/2015PEPI..249...43K} {249, 43}
       
       \bibitem[\protect\citeauthoryear{{Kong}, {Zhang}  \& {Schubert}}{{Kong}
         et~al.}{2016}]{kzs16}
       {Kong} D.,  {Zhang} K.,   {Schubert} G.,  2016, \mn@doi [\apj]
         {10.3847/0004-637X/826/2/127}, \href
         {https://ui.adsabs.harvard.edu/abs/2016ApJ...826..127K} {826, 127}
       
       \bibitem[\protect\citeauthoryear{{Kovetz}}{{Kovetz}}{1968}]{kov68}
       {Kovetz} A.,  1968, \mn@doi [\apj] {10.1086/149820}, \href
         {https://ui.adsabs.harvard.edu/abs/1968ApJ...154..999K} {154, 999}
       
       \bibitem[\protect\citeauthoryear{{Lanzano}}{{Lanzano}}{1962}]{lan62}
       {Lanzano} P.,  1962, \mn@doi [\icarus] {10.1016/0019-1035(62)90012-X}, \href
         {https://ui.adsabs.harvard.edu/abs/1962Icar....1..121L} {1, 121}
       
       \bibitem[\protect\citeauthoryear{{Lanzano}}{{Lanzano}}{1974}]{lan74}
       {Lanzano} P.,  1974, \mn@doi [\apss] {10.1007/BF00642721}, \href
         {https://ui.adsabs.harvard.edu/abs/1974Ap&SS..29..161L} {29, 161}
       
       \bibitem[\protect\citeauthoryear{Maclaurin}{Maclaurin}{1742}]{maclaurin42}
       Maclaurin C.,  1742, A Treatise of Fluxions. In Two Books. 1,
       T.W. and T. Ruddimans, Edinburgh
       
       \bibitem[\protect\citeauthoryear{{Maeder}}{{Maeder}}{2009}]{maeder2009}
       {Maeder} A.,  2009, {Physics, Formation and Evolution of Rotating Stars}, Springer Berlin, Heidelberg
         \mn@doi{10.1007/978-3-540-76949-1.
       }
       
       \bibitem[\protect\citeauthoryear{{Marchenko}}{{Marchenko}}{2000}]{mar00}
       {Marchenko} A.~N.,  2000, \mn@doi [Astronomical School's Report]
         {10.18372/2411-6602.01.1034}, \href
         {https://ui.adsabs.harvard.edu/abs/2000AstSR...1a..34M} {1, 34}
       
       \bibitem[\protect\citeauthoryear{{Miguel} \& {Vazan}}{{Miguel} \&
         {Vazan}}{2023}]{mv23}
       {Miguel} Y.,  {Vazan} A.,  2023, \mn@doi [Remote Sensing] {10.3390/rs15030681},
         \href {https://ui.adsabs.harvard.edu/abs/2023RemS...15..681M} {15, 681}
       
       \bibitem[\protect\citeauthoryear{{Nettelmann} et~al.,}{{Nettelmann}
         et~al.}{2021}]{net21}
       {Nettelmann} N.,  et~al., 2021, \mn@doi [The Planetary Science Journal]
         {10.3847/PSJ/ac390a}, \href
         {https://ui.adsabs.harvard.edu/abs/2021PSJ.....2..241N} {2, 241}
       
       \bibitem[\protect\citeauthoryear{Poincar{\'e}}{Poincar{\'e}}{1888}]{poincare88}
       Poincar{\'e} H.,  1888, Comptes-rendus des s{\'e}ances de l'Acad{\'e}mie des
         sciences, 106, 1571
       
       \bibitem[\protect\citeauthoryear{Ragazzo}{Ragazzo}{2020}]{ragazzo2020}
       Ragazzo C.,  2020, \mn@doi [São Paulo Journal of Mathematical Sciences]
         {10.1007/s40863-019-00162-3}, 14, 1
       
       \bibitem[\protect\citeauthoryear{{Rambaux}, {Chambat}  \&
         {Castillo-Rogez}}{{Rambaux} et~al.}{2015}]{rcc15}
       {Rambaux} N.,  {Chambat} F.,   {Castillo-Rogez} J.~C.,  2015, \mn@doi [\aap]
         {10.1051/0004-6361/201527005}, \href
         {https://ui.adsabs.harvard.edu/abs/2015A&A...584A.127R} {584, A127}
       
       \bibitem[\protect\citeauthoryear{{Roberts}}{{Roberts}}{1963}]{rob63}
       {Roberts} P.~H.,  1963, \mn@doi [\apj] {10.1086/147687}, \href
         {https://ui.adsabs.harvard.edu/abs/1963ApJ...138..809R} {138, 809}
       
       \bibitem[\protect\citeauthoryear{{Staelen}}{{Staelen}}{2022}]{sta22}
       {Staelen} C.,  2022, {Figures d’{\'e}quilibre à deux couches
         sph{\'e}ro{\"i}dales : Caract{\'e}risation des solutions avec rotation
         différentielle}, M.Sc. Dissertation, Universit{\'e} de Bordeaux
       
       \bibitem[\protect\citeauthoryear{{Tassoul}}{{Tassoul}}{1978}]{tassoul78}
       {Tassoul} J.-L.,  1978, {Theory of rotating stars}, Princeton University Press, Princeton
       
       \bibitem[\protect\citeauthoryear{Tisserand}{Tisserand}{1891}]{tisserand91}
       Tisserand F.,  1891, Trait{\'e} de m{\'e}canique c{\'e}leste - II. Th{\'e}orie
         de la figure des corps c{\'e}lestes et de leur mouvement de rotation.
       Gauthier-Villars et fils, Paris
       
       \bibitem[\protect\citeauthoryear{{Tricarico}}{{Tricarico}}{2014}]{tri14}
       {Tricarico} P.,  2014, \mn@doi [\apj] {10.1088/0004-637X/782/2/99}, \href
         {https://ui.adsabs.harvard.edu/abs/2014ApJ...782...99T} {782, 99}
       
       \bibitem[\protect\citeauthoryear{V\'eronet}{V\'eronet}{1912}]{veronet12}
       V\'eronet A.,  1912, Journal de math{\'e}matiques pures et appliqu{\'e}es 6e
         s{\'e}rie, 8, 331
       
       \bibitem[\protect\citeauthoryear{Volterra}{Volterra}{1903}]{1903volterra}
       Volterra V.,  1903, \mn@doi [Acta Mathematica] {10.1007/BF02421298}, 27, 105
       
       \bibitem[\protect\citeauthoryear{{Zahn}}{{Zahn}}{1992}]{zahn92}
       {Zahn} J.~P.,  1992, \aap, \href
         {https://ui.adsabs.harvard.edu/abs/1992A&A...265..115Z} {265, 115}
       
       \bibitem[\protect\citeauthoryear{{Zharkov} \& {Trubitsyn}}{{Zharkov} \&
         {Trubitsyn}}{1970}]{zt70}
       {Zharkov} V.~N.,  {Trubitsyn} V.~P.,  1970, \sovast, \href
         {https://ui.adsabs.harvard.edu/abs/1970SvA....13..981Z} {13, 981}
       
       \makeatother
       \end{thebibliography}

\appendix

\section{Kernel functions}\label{sec:kappa_chi_mu}

Let us write explicitely the kernel functions of the integrals of the main equations of the present work. 

$\forall \varpi\in{]0,1]}, \forall \varpi'\in{[0,\varpi[}$, we have
\begin{align}
       \kappa^{\rm in}(\varpi',\varpi) = &\frac{\bare(\varpi')}{\epsilon^3(\varpi')}\bigg\{\left[1-2\frac{\varpi'^2\epsilon^2(\varpi')}{\varpi^2}\right]\arcsin\left(\frac{\varpi'\epsilon(\varpi')}{\varpi}\right)\notag\\
       &-\frac{\varpi'\epsilon(\varpi')}{\varpi}\left(2\bare(\varpi)+\sqrt{1-\frac{\varpi'^2\epsilon^2(\varpi')}{\varpi^2}}\right)\notag\\
       &+2\arcsin\left(\frac{\varpi'\epsilon(\varpi')}{\varpi\sqrt{1+c(\varpi',\varpi)}}\right)\big[1+c(\varpi',\varpi)\big]\bigg\},\label{eq:kin_def}
\end{align}

$\forall \varpi\in[0,1], \forall \varpi'\in[a,1]$, we have
\begin{align}
       \kappa^{\rm out}(\varpi',\varpi) = &\left[3-2\epsilon^2(\varpi)\right]\notag\\
       &\times\left[\frac{\bare(\varpi')}{\epsilon^3(\varpi')}\arcsin\big(\epsilon(\varpi')\big)-\frac{1}{\epsilon^2(\varpi')}\right]+1 \label{eq:kout_def}
\end{align}

One could be worried by the multiple divergences in $\kappa^{\rm in}$ at $\varpi=0$. Yet, in this case, we see that the first integral in \eqref{eq:drhoom2_nsfoe} vanishes and the divergences are then never taken into account in the calculations.

The property $\kappa^{\rm in}(\varpi,\varpi) = \kappa^{\rm out}(\varpi,\varpi)$ is easily proven by remembering $c(\varpi,\varpi)=0$.

The derivative functions of $\kappa^{\rm in}$ and $\kappa^{\rm out}$ written in Eqs. \eqref{eq:dkin} and \eqref{eq:dkout} are given by
\begin{align}
       \chi(\varpi',\varpi) = &\frac{\varpi'^2\bare(\varpi')}{\varpi^3\epsilon(\varpi')}\bigg[\arcsin\left(\frac{\varpi'\epsilon(\varpi')}{\varpi}\right)\notag\\
       &-\arcsin\left(\frac{\varpi'\epsilon(\varpi')}{\varpi\sqrt{1+c(\varpi',\varpi)}}\right)\bigg],\label{eq:chi_def}\\
       \mu(\varpi',\varpi) =& \frac{\bare(\varpi')}{\epsilon^3(\varpi')}\left[\arcsin\left(\frac{\varpi'\epsilon(\varpi')}{\varpi\sqrt{1+c(\varpi',\varpi)}}\right)-\frac{\varpi'\epsilon(\varpi')}{\varpi\bare(\varpi)}\right],\label{eq:mu_def}\\
       \nu(\varpi') =& \frac{\bare(\varpi')}{\epsilon^3(\varpi')}\arcsin(\epsilon(\varpi'))-\frac{1}{\epsilon^2(\varpi')},\label{eq:nu_def}
\end{align}
where $\chi$ and $\mu$ are defined for $\varpi\in{]0,1]}, \varpi'\in{[0,\varpi[}$ and $\nu$ is defined for $\varpi'\in[0,1]$.

Once again, we can easily prove that $\chi(\varpi,\varpi)=0$ and $\mu(\varpi,\varpi) = \nu(\varpi,\varpi)$, leading to the continuity in $\varpi'=\varpi$ of the derivative of the $\kappa$-functions.

\section{Roberts' equation}\label{sec:rob}

Equation (3.23) of \citet{rob63} reads
\begin{align}
       \frac{\du\epsilon^2}{\du \varpi}&\bigg\{\frac{16\uppi}{45}\hatomega^2\varpi^3\bare(\varpi)-\frac{4}{15}\check M(\varpi) - \frac25\check D(\varpi) \notag\\
       &+\frac{3-2\epsilon^2(\varpi)}{2\epsilon^5(\varpi)}\bare(\varpi)\bigg\{\left[3-2\epsilon^2(\varpi)\right]\arcsin\big(\epsilon(\varpi)\big)\notag\\
       &-3\epsilon(\varpi)\bare(\varpi)\bigg\}\check D(\varpi)\bigg\}\notag\\
       =&\frac{3-2\epsilon^2(\varpi)}{\varpi\epsilon^3(\varpi)}\bare(\varpi)\bigg\{\left[3-2\epsilon^2(\varpi)\right]\arcsin\big(\epsilon(\varpi)\big)\notag\\
       &-3\epsilon(\varpi)\bare(\varpi)\bigg\}\check D(\varpi),\label{eq:roberts}
\end{align}
where
\begin{equation}
       \check M(\varpi) = 4\uppi \int_0^\varpi\du \varpi' \frac{\varpi'\hatrho(\varpi')}{\bare(\varpi')}\left[1-\epsilon^2(\varpi)-\frac{\varpi'}{6}\frac{\du\epsilon^2}{\du \varpi'}\right]
\end{equation}
and 
\begin{align}
       \check D(\varpi) = &4\uppi \int_0^\varpi \du \varpi' \frac{\varpi'\hatrho(\varpi')}{\bare(\varpi')}\Bigg\{\left[1-\epsilon^2(\varpi')\right]\left[1-\frac53\frac{\varpi'^2\epsilon^2(\varpi')}{\varpi^2\epsilon^2(\varpi)}\right]\notag\\
       &- \frac{\varpi'}{6}\frac{\du\epsilon^2}{\du \varpi'}\left[1+\frac{\varpi'^2[2-3\epsilon^2(\varpi')]}{\varpi^2\epsilon^2(\varpi)}\right]\Bigg\}.
\end{align}

\section{On the volume integrals}\label{sec:intvol}

The quantities calculated in Tabs. \ref{tab:configA}, \ref{tab:configB}, \ref{tab:configC}, \ref{tab:configD} write
\begin{equation}\label{eq:volint}
       \left\{
              \begin{aligned}
                     &M = \int \du V \rho(\varpi),\\
                     &V = \int \du V \left(=\frac43\uppi R_{\rm e}^3\bare(1)\right),\\
                     &J = \int \du V \rho(\varpi) \Omega(\varpi) R^2,\\
                     &W = \frac12\int \du V \rho(\varpi) \varPsi(\varpi,\theta),\\
                     &U = 3\int \du V p(\varpi), \\
                     &T = \frac12\int \du V \rho(\varpi) \Omega^2(\varpi) R^2.
              \end{aligned}
       \right.
\end{equation}

So we need to express the volume element $\du V$ and the cylindrical radius $R$ as functions of the spherical polar angle $\theta$ and $\varpi$. Along an isopycnic surface, the spherical radius $r$ reads
\begin{equation}
       \frac{r}{R_{\rm e}} = \frac{\varpi\sqrt{1-\epsilon^2(\varpi)}}{\sqrt{1-\epsilon^2(\varpi)\sin^2(\theta)}}.
\end{equation}

The Jacobian matrix ${\mathfrak J}$ of the transformation from cartesian coordinates to an isopycnic coordinate system $(\varpi,\theta,\varphi)$, where $\varphi$ is the spherical azimutal angle, reads
\begin{equation}
       {\mathfrak J} = \begin{pmatrix}
              \frac{\partial r}{\partial \varpi}\sin(\theta)\cos(\varphi) & \frac{r\cos(\theta)\cos(\varphi)}{1-\epsilon^2(\varpi)\sin^2(\theta)} & -r\sin(\theta)\sin(\varphi) \\
              \frac{\partial r}{\partial \varpi}\sin(\theta)\sin(\varphi) & \frac{r\cos(\theta)\sin(\varphi)}{1-\epsilon^2(\varpi)\sin^2(\theta)} & r\sin(\theta)\cos(\varphi) \\
              \frac{\partial r}{\partial \varpi}\cos(\theta) & -\frac{r\sin(\theta)[1-\epsilon^2(\varpi)]}{1-\epsilon^2(\varpi)\sin^2(\theta)}  & 0 \\
       \end{pmatrix},
\end{equation}
where $(\partial r/\partial \varpi)$ reads 
\begin{align}
       \frac{1}{R_{\rm e}}\frac{\partial r}{\partial \varpi} = &\frac{\sqrt{1-\epsilon^2(\varpi)}}{\sqrt{1-\epsilon^2(\varpi)\sin^2(\theta)}} \notag\\
       &- \frac{\varpi}{2}\frac{\du\epsilon^2}{\du \varpi}\frac{\cos^2(\theta)}{\sqrt{1-\epsilon^2(\varpi)}[1-\epsilon^2(\varpi)\sin^2(\theta)]^{3/2}}. 
\end{align}

So, the volume element is then given by
\begin{align}
       \frac{\du V}{R_{\rm e}^3} =& \frac{\det({\mathfrak J})}{R_{\rm e}^3}\du\varpi\du\theta\du\varphi \\
       =& \frac{\left[1-\epsilon^2(\varpi)\right]\left[1-\epsilon^2(\varpi)\sin^2(\theta)\right]-\dfrac{\varpi}{2}\dfrac{\du\epsilon^2}{\du \varpi}\cos^2(\theta)}{[1-\epsilon^2(\varpi)\sin^2(\theta)]^{5/2}} \notag\\
       &\times\varpi^2\sqrt{1-\epsilon^2(\varpi)}\sin(\theta) \du\varpi \du\theta \du\varphi.
\end{align}

The cylindrical radius is given by $R=r\sin(\theta)$, namely
\begin{equation}
       \frac{R}{R_{\rm e}} = \frac{\varpi\sqrt{1-\epsilon^2(\varpi)}\sin(\theta)}{\sqrt{1-\epsilon^2(\varpi)\sin^2(\theta)}}.
\end{equation}

All the integrals of \eqref{eq:volint} can now be computed numerically ({\it via} a trapezoidal rule for instance).

For the rotation rate, as it is not exactly a constant due to the spheroidal approximation, we can obtain an mean value by integrating over the mass, namely
\begin{equation}\label{eq:om2_mean}
       \big\langle\Omega\big\rangle = \frac{\int \du V \rho(\varpi) \Omega(\varpi)R^2}{\int \du V \rho(\varpi)R^2}. \left(=\frac{J}{I}\right)
\end{equation}

\label{lastpage}
\end{document}